\documentclass[]{article}
\usepackage[top=1in, bottom=1in, left=1in, right=1in]{geometry}

\usepackage[doublespacing]{setspace}
\usepackage{subfigure}
\usepackage[numbers,sort&compress]{natbib}

\bibpunct[, ]{[}{]}{,}{n}{,}{,}

\usepackage[normalem]{ulem} 

\usepackage{amsmath}
\usepackage{amssymb}
\numberwithin{equation}{section} 
\usepackage{float}
\usepackage{placeins}
\usepackage{comment}
\usepackage[affil-it]{authblk}
\date{\today}

\newcommand{\pdv}[2]{\frac{\partial #1}{\partial #2}}

\usepackage{tikz}
\usetikzlibrary{plotmarks}

\NewDocumentCommand{\Plot}{ m O{} }{%
    (%
    \tikz[baseline=-0.6ex,inner sep=0pt, outer xsep=0pt] {%
        \draw[line width=1pt,text height = \textheight,#1] plot coordinates {(0,0)} -- plot[#2,mark options={solid}, mark size=2pt] coordinates {(2mm,0)} -- plot coordinates {(4mm,0)};%
    }%
    )%
}

\definecolor{firebrick}{rgb}{0.7, 0.13, 0.13}

\title{Fully Turbulent Wakes at Low Reynolds Numbers: the Case of the Thin Flat Plate}

\author[a]{Isaac T. Rosin}
\author[a]{Melanie S. Chapman}
\author[b]{Bartosz Protas}
\author[a]{Robert J. Martinuzzi\thanks{For correspondence: rmartinu@ucalgary.ca}}

\affil[a]{Department of Mechanical \& Manufacturing Engineering, University of Calgary, Calgary, Alberta, Canada}
\affil[b]{Department of Mathematics \& Statistics, McMaster University, Hamilton, Ontario, Canada}

\begin{document}

\maketitle

\begin{abstract}
    We consider the wake flow past a thin two-dimensional flat plate normal to the uniform stream and demonstrate that this flow is turbulent already at a relatively low Reynolds number of $Re = 400$. This is achieved by performing a careful comparison of the results of a DNS of this flow with experimental measurements of wake flows in the same geometric configuration at the Reynolds numbers of $Re=12500, 19700$. This comparison reveals that the distribution of several key quantities, including the mean velocity, Reynolds stresses and different effects contributing to the transport of the turbulent kinetic energy, are, up to measurement uncertainty, the same in these flows. Moreover, the wake flow at $Re = 400$ also features energy spectra characteristic of turbulent flows with intermittency detected in the distributions of the fluctuating strain and rotation rates. In contrast, these features are absent from the results of the DNS of the wake flow at $Re = 150$  where the distribution of the key quantities is also fundamentally different. These results show that the path to transition to turbulence in the wake past a thin flat plate is different from that in the wakes of canonical (i.e., circular or square) cylinders. We also identify possible physical mechanisms that may be responsible for these differences.
\end{abstract}

\begin{center}
    \textbf{Keywords:} Turbulent wake; thin flat plate; low Reynolds number; Direct Numerical Simulation
\end{center}
    


\section{Introduction}

Despite its simple geometry, the numerical simulation of the nominally two-dimensional (2D) flow past a thin flat plate of infinite span normal to a uniform stream has proven challenging~\cite{najjarbalachandar1998,hemmati2016,hemmati2018}.
While 2D in the time-averaged sense, the instantaneous wake is intrinsically three-dimensional (3D) even at low Reynolds numbers ($Re \gtrsim 105-110$ \cite{choiyang2014, thompson2006, freeman2023}), with $Re$ based on the flow-normal plate width $d$, the free-stream velocity $U_\infty$ and the kinematic viscosity $\nu$.
Experimental studies indicate that integral parameters, such as force coefficients and the non-dimensional vortex shedding frequency (the Strouhal number $St$) depend weakly on $Re$ in the 3D regime even at low $Re$~\cite{roshko1954,polhamus1984,song2020,braun2020}.
In contrast, computational fluid dynamics (CFD) studies do not find a consensus. In particular for $Re < 1000$, CFD results have reported very different wake dynamics, {\it cf.}~the reviews~\cite{hemmati2016,hemmati2018}.
A comparison of these different studies raises the possibility that the wake is already fully turbulent immediately above the onset of 3D instabilities in contrast to canonical cylinder wakes. Here ``fully turbulent" is understood to mean that the flow is spatio-temporally complex with velocity fluctuations exhibiting broadband spectra with an inertial range consistent with Kolmogorov's $-5/3$ law \cite{frisch1995turbulence}, their statistical moments slow functions of $Re$ and the probability density functions (PDFs) of the invariants of the fluctuating deformation and rotation tensors characteristically non-Gaussian \cite{buaria2019}.

 The laminar-turbulent wake transition over a very short $Re$ range would be a very different behaviour compared to the canonical circular and square cylinder wakes. Here we define a cylinder as an object that is symmetric about a flow centre plane, is geometrically homogeneous and  sufficiently long in the span so that the mean flow field is 2D. Wakes behind circular and square cylinders appear to undergo several stages of transition, typically over $200 < Re < 10000$, where the wake exhibits strong $Re$-dependence and involves regimes where velocity fluctuations do not manifest turbulent behaviour \cite{williamson1996,ryan2005,mavriplis2017}.
The objective of the present communication is to demonstrate that the thin flat plate wake at $Re=400$ exhibits a turbulent behaviour qualitatively similar to flows with $Re>1000$. These observations are supported by detailed comparisons to experimental data at $Re=12500$ and $Re = 19700$.

Below we provide an overview of the $Re$-dependent laminar-turbulent transition processes occurring in the wakes behind circular and square cylinders and thin flat plates. The influence of the streamwise thickness $b$ to $d$ aspect ratio (AR) and of the presence of sharp corners on these processes is discussed. Then, the specific research objective of the present study is formulated.

The review of the circular cylinder wake by Williamson \cite{williamson1996} describes seven $Re$-dependent flow regimes. The end of the steady-flow regime is marked by two-dimensional instabilities starting with the Hopf bifurcation at $Re \approx 48$ and the emergence of periodic von K\'{a}rm\'{a}n vortex shedding of spanwise vortices (rollers). Five regimes exist above the appearance of 3D instabilities at $Re \approx 180$, culminating in the boundary layer on the cylinder surface transitioning to turbulence at $Re\approx 10^5$. As $Re$ increases above the onset of 3D instabilities, mode A, then mode B and mode C instabilities emerge~\cite{ryan2005,thompson2006,choiyang2014}, after which the flow is in a stable 3D quasi-periodic vortex-shedding state. Mode A is a spanwise shear layer instability appearing at $Re\approx180$. It is characterized by a spanwise wavelength of about 4 cylinder diameters and is often associated with instability of the vortex core, resulting in distortion and dislocations of the rollers. Mode B appears for $Re\gtrsim220$ and consists of counter-rotating streamwise vortices, known as ``braids" or ``ribs", winding between the rollers and spaced roughly one diameter apart spanwise. Mode C appears for $170 \lesssim Re \lesssim 270$ and has a spanwise wavelength of about two diameters~\cite{zhang1995}. It is manifested as spanwise counter-rotating vortex pairs
linked to period doubling (subharmonic of the fundamental shedding frequency)~\cite{sheard2005}.

Laminar-turbulent transition occurs after the appearance of the aforementioned instabilities. Initially, it occurs downstream of the base flow region, nominally the formation region of the von K\'{a}rm\'{a}n vortices. It progressively moves closer to the cylinder as $Re$ increases, such that the wake dynamics integral values (e.g. drag) change rapidly with increasing $Re$. For $10^4 \lesssim Re \lesssim 10^5$, transition occurs immediately downstream of where the flow separates from the cylinder. 
%
Changes in the wake dynamics (shedding frequency, moments of the fluctuating velocity or invariants of the deformation and rotation tensors) and integral quantities exhibit weak $Re$-dependence. In this regime, the wake is then said to be fully turbulent. The laminar-turbulent transition of the boundary layer ($Re \sim 10^5$) results in a sudden delay in the boundary layer separation and a rapid decrease in the drag coefficient \cite{singh2005} ({\it i.e.}, the drag crisis).


Studies on square cylinder wakes suggest that the transition process and path are similar to those for circular cylinder wakes~\cite{ryan2005,thompson2006,mavriplis2017}.
Analogous types of instabilities appear in the same sequence, although the critical $Re$ for the onset of 3D instabilities differ and the fully turbulent wake ($Re$-insensitive regime) is already  observed above $Re \approx 6000$.
These differences are ostensibly due to the fixed separation point at the sharp leading edge, which also explains the absence of a drag crisis.

For elliptical and rectangular cylinders with aspect ratio $b/d$, the  types and sequences of instabilities observed, as well as the process leading to transition, are similar to the canonical cases for $0.5 \lesssim b/d \lesssim 3$. For elongated cylinders, $b/d > 3$, the flow reattaches on the obstacles sides and the wakes evolve differently. 

For shorter cylinders, $b/d < 0.5-0.6$, the wake dynamics develop differently from the canonical cases~\cite{braun2020,choiyang2014,thompson2014,ongyin2022}.  For $b/d < 0.1-0.2$, the elliptical and rectangular cylinder wakes are very similar and are insensitive to rounded or sharp edges~\cite{choiyang2014,thompson2014}. Cylinders in the latter aspect ratio range can be considered thin flat plates.

Floquet stability analysis for elliptic~\cite{blackburn2010,thompson2014} and rectangular~\cite{choiyang2014} cylinder wakes indicates that mode B and C instabilities cease to be observed below $b/d\approx 0.4$ while mode A is not observed for $b/d \lesssim 0.1$. These are replaced by quasi-periodic (QP) spanwise instabilities and vortex core instabilities (similar to mode A). The critical $Re$ for these instabilities is between 100 and 150.  The QP modes are most unstable, modifying the wake of thin flat plates.

Whereas earlier work~\cite{tian2014,hemmati2018} suggests that wakes of thin flat plates vary slowly with $Re$ above $1200$, the reported wake behaviour differs significantly quantitatively and qualitatively when comparing numerical simulations for $Re < 1000$.  While some studies suggest a strong $Re$-dependence~\cite{joshi1993,najjarbalachandar1998,narasimhamurthy2009}, others suggest that the wake varies little above $Re=500$~\cite{choiyang2014,ongyin2022}. It is noted, however, that the simulations in the studies~\cite{choiyang2014,ongyin2022} show different wake structures and dynamics. This disagreement has not been satisfactorily explained. Hence, in this communication, two independent simulations are conducted using different discretization schemes, namely, the specteral-element and finite-volume methods.  

The idea that the thin flat plate wake may transition to turbulence for $Re<1000$ following a different sequence of instabilities would demonstrate that this geometry produces a fundamentally different wake to those of classical cylinders. The main goal of this communication is to provide and discuss evidence that this flow is indeed fully turbulent already at $Re = 400$ based on a comparison of numerical and experimental data. 
Simulations at $Re=150$ are also presented to provide a comparison to a case which is not fully turbulent but exhibits a wake~\cite{ongyin2022} with 3D instabilities. The structure of the paper is as follows: Section \ref{sec:methods} details the computational and experimental methodology with which the data is obtained, Section \ref{sec:results} provides a thorough characterization of the flows investigated with an emphasis on their turbulent properties, whereas concluding remarks are deferred to Section \ref{sec:conclusion}.

\FloatBarrier

\section{Methodology}\label{sec:methods}

In this section, we first define the flow domain and state the governing equations together with the boundary conditions. Then, we outline and validate the numerical approaches used to solve the problem. Finally, we present experimental data that will be used for comparison with the results of the numerical simulations.

\subsection{Flow Domain}

Throughout this work we use the Cartesian coordinate system $\{x,y,z\}$ with the origin $O$ placed at the centre of the leeward surface of the thin flat plate $S$. The position vector is denoted $\mathbf{x} = [x, y, z]$. 
While the idealized  flow domain is externally unbounded, {\it i.e.}, has the form $\mathbb{R}^3 \backslash S$, it must be truncated when defining the problem to be solved computationally. Our computational domain is shown in the $x$-$y$ plane in  Figure \ref{fig:2D_ref_diagram}, whereas Figure \ref{fig:3D_ref_diagram} offers an isometric view.  The unit vector normal to the domain boundary $\Gamma$ and pointing outside is denoted $\mathbf{n}$.

\begin{figure}
    \centering 
    \tikzset{every picture/.style={line width=0.75pt}} 

\begin{tikzpicture}[x=0.75pt,y=0.75pt,yscale=-0.75,xscale=0.75]

\draw  [line width=1.5]  (54,38) -- (639,38) -- (639,398) -- (54,398) -- cycle ;
\draw  [line width=1.5]  (201.13,194.28) -- (205.41,194.28) -- (205.41,241.59) -- (201.13,241.59) -- cycle ;
\draw    (59.33,68.17) -- (87.33,68.17) ;
\draw [shift={(89.33,68.17)}, rotate = 180] [color={rgb, 255:red, 0; green, 0; blue, 0 }  ][line width=0.75]    (10.93,-3.29) .. controls (6.95,-1.4) and (3.31,-0.3) .. (0,0) .. controls (3.31,0.3) and (6.95,1.4) .. (10.93,3.29)   ;

\draw  [dash pattern={on 0.84pt off 2.51pt}]  (54.33,218) -- (213.15,218) -- (637,218.01) ;
\draw [shift={(639,218.01)}, rotate = 180] [fill={rgb, 255:red, 0; green, 0; blue, 0 }  ][line width=0.08]  [draw opacity=0] (12,-3) -- (0,0) -- (12,3) -- cycle    ;
\draw  [dash pattern={on 0.84pt off 2.51pt}]  (206,397.52) -- (206,218.38) -- (206,39.52) ;
\draw [shift={(206,37.52)}, rotate = 90] [fill={rgb, 255:red, 0; green, 0; blue, 0 }  ][line width=0.08]  [draw opacity=0] (12,-3) -- (0,0) -- (12,3) -- cycle    ;
\draw  (73.43,378.56) -- (103.43,378.56)(73.43,348.56) -- (73.43,378.56) -- cycle (96.43,373.56) -- (103.43,378.56) -- (96.43,383.56) (68.43,355.56) -- (73.43,348.56) -- (78.43,355.56)  ;

\draw    (195,194) -- (195,242) ;
\draw [shift={(195,242)}, rotate = 270] [color={rgb, 255:red, 0; green, 0; blue, 0 }  ][line width=0.75]    (0,2.24) -- (0,-2.24)   ;
\draw [shift={(195,194)}, rotate = 270] [color={rgb, 255:red, 0; green, 0; blue, 0 }  ][line width=0.75]    (0,2.24) -- (0,-2.24)   ;
\draw    (43,38) -- (43,218) ;
\draw [shift={(43,218)}, rotate = 270] [color={rgb, 255:red, 0; green, 0; blue, 0 }  ][line width=0.75]    (0,5.59) -- (0,-5.59)   ;
\draw [shift={(43,38)}, rotate = 270] [color={rgb, 255:red, 0; green, 0; blue, 0 }  ][line width=0.75]    (0,5.59) -- (0,-5.59)   ;
\draw    (54.5,406) -- (206,406) ;
\draw [shift={(206,406)}, rotate = 180] [color={rgb, 255:red, 0; green, 0; blue, 0 }  ][line width=0.75]    (0,5.59) -- (0,-5.59)   ;
\draw [shift={(54.5,406)}, rotate = 180] [color={rgb, 255:red, 0; green, 0; blue, 0 }  ][line width=0.75]    (0,5.59) -- (0,-5.59)   ;
\draw    (206,406) -- (639,406) ;
\draw [shift={(639,406)}, rotate = 180] [color={rgb, 255:red, 0; green, 0; blue, 0 }  ][line width=0.75]    (0,5.59) -- (0,-5.59)   ;
\draw    (43,218) -- (43,397.75) ;
\draw [shift={(43,397.75)}, rotate = 270] [color={rgb, 255:red, 0; green, 0; blue, 0 }  ][line width=0.75]    (0,5.59) -- (0,-5.59)   ;

\draw (58.33,42.57) node [anchor=north west][inner sep=0.75pt]    {$U_{\infty }$};
\draw (622,223) node [anchor=north west][inner sep=0.75pt]    {$x$};
\draw (187.33,41) node [anchor=north west][inner sep=0.75pt]    {$y$};
\draw (91,384) node [anchor=north west][inner sep=0.75pt]    {$u$};
\draw (55.77,341.62) node [anchor=north west][inner sep=0.75pt]    {$v$};
\draw (179,209.27) node [anchor=north west][inner sep=0.75pt]    {$d$};
\draw (1,121.44) node [anchor=north west][inner sep=0.75pt]    {$7.5d$};
\draw (117,407.78) node [anchor=north west][inner sep=0.75pt]    {$10d$};
\draw (411,408.78) node [anchor=north west][inner sep=0.75pt]    {$30d$};
\draw (1,299.44) node [anchor=north west][inner sep=0.75pt]    {$7.5d$};

\end{tikzpicture}
    \caption{Schematic of the computational domain in the $x$--$y$ plane.}
    \label{fig:2D_ref_diagram}
\end{figure}
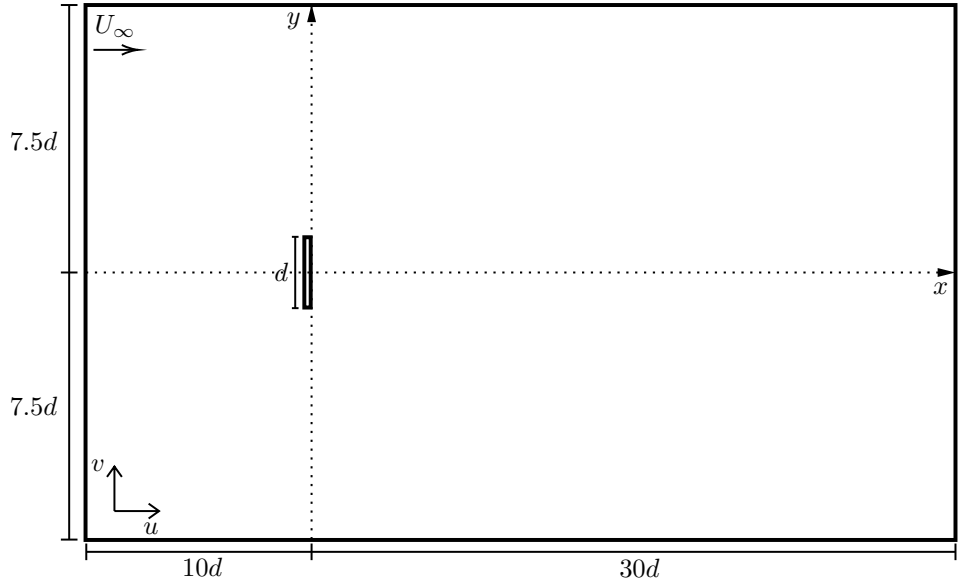
\begin{figure}
    \centering
    \tikzset{every picture/.style={line width=0.75pt}} 

\begin{tikzpicture}[x=0.75pt,y=0.75pt,yscale=-1.5,xscale=1.5]

\draw    (0,90) -- (30,130) ;
\draw    (0,230) -- (30,270) ;
\draw    (70,150) -- (70,170) ;
\draw    (0,230) -- (260,230) ;
\draw    (70,170) -- (100,210) ;
\draw    (70,150) -- (100,190) ;
\draw    (100,190) -- (100,210) ;
\draw    (105,190) -- (105,210) ;
\draw    (100,210) -- (105,210) ;
\draw    (100,190) -- (105,190) ;
\draw    (70,150) -- (75,150) ;
\draw    (75,150) -- (105,190) ;
\draw [line width=0.75]    (66.71,171.86) -- (95.86,211.86) ;
\draw [shift={(95.86,211.86)}, rotate = 233.92] [color={rgb, 255:red, 0; green, 0; blue, 0 }  ][line width=0.75]    (0,2.24) -- (0,-2.24)   ;
\draw [shift={(66.71,171.86)}, rotate = 233.92] [color={rgb, 255:red, 0; green, 0; blue, 0 }  ][line width=0.75]    (0,2.24) -- (0,-2.24)   ;

\draw    (96.8,173.34) -- (118.8,173.34) ;
\draw [shift={(121.8,173.34)}, rotate = 180] [fill={rgb, 255:red, 0; green, 0; blue, 0 }  ][line width=0.08]  [draw opacity=0] (3.57,-1.72) -- (0,0) -- (3.57,1.72) -- cycle    ;
\draw    (96.8,173.34) -- (110,190.94) ;
\draw [shift={(111.8,193.34)}, rotate = 233.13] [fill={rgb, 255:red, 0; green, 0; blue, 0 }  ][line width=0.08]  [draw opacity=0] (3.57,-1.72) -- (0,0) -- (3.57,1.72) -- cycle    ;
\draw    (96.8,173.34) -- (96.8,151.34) ;
\draw [shift={(96.8,148.34)}, rotate = 90] [fill={rgb, 255:red, 0; green, 0; blue, 0 }  ][line width=0.08]  [draw opacity=0] (3.57,-1.72) -- (0,0) -- (3.57,1.72) -- cycle    ;

\draw    (30,270) -- (290,270) ;
\draw    (30,130) -- (290,130) ;
\draw    (0,90) -- (260,90) ;
\draw    (260,90) -- (290,130) ;
\draw    (260,230) -- (290,270) ;
\draw    (0,90) -- (0,230) ;
\draw    (30,130) -- (30,270) ;
\draw    (260,90) -- (260,230) ;
\draw    (290,130) -- (290,270) ;
\draw    (58.43,148.14) -- (65.22,149.74) ;
\draw [shift={(68.14,150.43)}, rotate = 193.24] [fill={rgb, 255:red, 0; green, 0; blue, 0 }  ][line width=0.08]  [draw opacity=0] (3.57,-1.72) -- (0,0) -- (3.57,1.72) -- cycle    ;
\draw    (85.86,147.29) -- (79.86,149.22) ;
\draw [shift={(77,150.14)}, rotate = 342.12] [fill={rgb, 255:red, 0; green, 0; blue, 0 }  ][line width=0.08]  [draw opacity=0] (3.57,-1.72) -- (0,0) -- (3.57,1.72) -- cycle    ;
\draw [line width=0.75]    (-4.49,231.46) -- (24.66,271.46) ;
\draw [shift={(24.66,271.46)}, rotate = 233.92] [color={rgb, 255:red, 0; green, 0; blue, 0 }  ][line width=0.75]    (0,2.24) -- (0,-2.24)   ;
\draw [shift={(-4.49,231.46)}, rotate = 233.92] [color={rgb, 255:red, 0; green, 0; blue, 0 }  ][line width=0.75]    (0,2.24) -- (0,-2.24)   ;

\draw (74,191) node [anchor=north west][inner sep=0.75pt]  [font=]  {$s$};
\draw (109.49,193) node [anchor=north west][inner sep=0.75pt]  [font=]  {$z$};
\draw (122,170) node [anchor=north west][inner sep=0.75pt]  [font=]  {$x$};
\draw (93,140.29) node [anchor=north west][inner sep=0.75pt]  [font=]  {$y$};
\draw (68,140) node [anchor=north west][inner sep=0.75pt]  [font=]  {$b$};
\draw (240,258) node [anchor=north west][inner sep=0.75pt]  [font=]  {Front $L_{xy}$};
\draw (218,218) node [anchor=north west][inner sep=0.75pt]  [font=]  {Back  $L_{xy}$};
\draw (-10,250) node [anchor=north west][inner sep=0.75pt]  [font=]  {$10d$};

\end{tikzpicture}
    \caption{Isometric view of the computational domain.}
    \label{fig:3D_ref_diagram}
\end{figure}
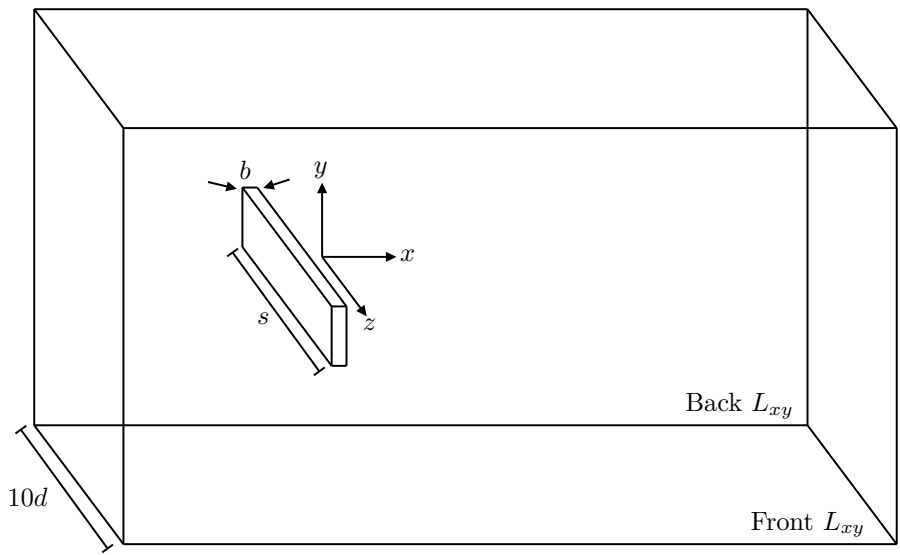

\subsection{Governing Equations and Boundary Conditions}

The flow is assumed to be governed by the incompressible Navier-Stokes equations stated here in a form nondimensionalized with respect to the free-stream velocity $U_{\infty}$, the fluid density $\rho$ and a characteristic length-scale given by the plate flow-normal width $d$
\begin{align}
    \pdv{\mathbf{u}}{t} + (\mathbf{u}\cdot\nabla)\mathbf{u} - \frac{1}{Re}\nabla^2\mathbf{u} & = - \nabla p,\label{eq:mom} \\
        \nabla\cdot\mathbf{u} & = 0, \label{eq:inc}
\end{align}
where $\mathbf{u} = [u, v, w]^T$ is the velocity vector, $p$ is pressure, and $t$ is time, whereas the Reynolds number $Re$ is defined as
\begin{gather}
    Re = \frac{U_\infty d}{\nu} .
\end{gather}

In the results section, statistical moments are presented following Reynolds decomposition into time average, or mean, and fluctuating velocity components
\[ [u,v,w]=[U,V,W]+[u',v',w']  \,\, {,}
\]
where the time and space arguments are implied. For higher moments, an overbar is used to indicate mean values, {\it e.g.} $\overline{u'^2}$. 

System \eqref{eq:mom}--\eqref{eq:inc} is solved subject to the following boundary conditions (see Figures \ref{fig:2D_ref_diagram} and \ref{fig:3D_ref_diagram} for the definitions of the different parts of the boundary):
\begin{itemize}
\item no-slip and no-through flow on the plate surface
\begin{equation}
    \mathbf{u} = \mathbf{0}, \qquad \qquad \mathbf{x} \in \partial S,
    \label{eq:BCS}
\end{equation}

\item  unperturbed free-stream at the inlet
\begin{equation}
    \mathbf{u} = [1,0,0] \, , \qquad \qquad \mathbf{x} \in \Gamma_{\text{i}} ,
\end{equation}

\item the condition proposed by Dong et al.~\cite{dongkarniadakis2014} at the outlet
\begin{equation}
    \mathbf{n}\cdot\nabla\mathbf{u} = Re\, p \, \mathbf{n} + \frac{Re}{4}|\mathbf{u}|^2\left(1 - \tanh{\frac{\mathbf{n}\cdot\mathbf{u}}{U_0\delta}}\right)\mathbf{n},
    \qquad \qquad \mathbf{x} \in \Gamma_{\text{o}},
\end{equation} 
where $U_0=1$ is the characteristic nondimensional velocity scale, and $\delta$ is a positive dimensionless constant chosen to be 0.05;

\item slip velocity on the lateral boundaries 
\begin{equation}
    \mathbf{n}\cdot\nabla\mathbf{u} = 0, \qquad \qquad \mathbf{x} \in \Gamma_{\text{l}},
\end{equation}

\item periodicity in the spanwise ($z$) direction
\begin{equation}
\left.
\begin{aligned}
    \mathbf{u}\big|_{z=-5} & = \mathbf{u}\big|_{z=5} \\
    \mathbf{n}\cdot\nabla\mathbf{u}\big|_{z=-5} & = 
    \mathbf{n}\cdot\nabla\mathbf{u}\big|_{z=5}
\end{aligned}
\quad \right\} \quad \text{for} \quad (x,y) \in L_{xy},
\label{eq:BCper}
\end{equation}
where $L_{xy} = [-10.05d,29.95d] \times [-7.5d,7.5d]$.
\end{itemize}

\subsection{Numerical Methods}

In order to perform direct numerical simulations (DNS) of the flows of interest,
system \eqref{eq:mom}--\eqref{eq:inc} subject to the boundary conditions \eqref{eq:BCS}--\eqref{eq:BCper} is solved using two distinct numerical techniques, namely, the spectral/$hp$ element method (SEM) implemented in the open source code {\tt Nektar++} \cite{nektar++2015} and the finite volume method (FVM) implemented in the open source code {\tt OpenFOAM} \cite{openfoam1998}. While the SEM simulations are used to produce the bulk of the DNS data analyzed in Section \ref{sec:results}, the computations performed with the FVM allow us to conduct an initial study of grid dependence and to cross-validate the SEM computations which is done in Section \ref{sub:conval}.

Our goal in this study is to conduct an accurate DNS of turbulent flows and then study certain fine properties of these flows. Therefore, our choice to rely on the SEM approach for most of the data is dictated by its higher numerical accuracy in terms of the spatial discretization. More precisely, being based on a high-order spectral-element approximation, this approach has a lower numerical diffusivity than the FVM, thus making it easier to accurately resolve all relevant scales of motion, down to the Kolmogorov scale.
Furthermore, SEM is able to achieve the same degree of accuracy and convergence in high-order statistics as FVM with fewer degrees of freedom~\cite{capuano2019,theobald2020}.

\subsection{Convergence and Validation}\label{sub:conval}

Mesh convergence is studied at $Re=400$ and is assessed based on the mean drag coefficient $\overline{C}_D$ and the Strouhal frequency $St$, defined as 
\begin{gather}
    \overline{C}_D = \frac{2\overline{F}_D}{\rho U_\infty^2 A},
    \label{eq:CD} \\
    St = \frac{fd}{U_\infty},   
    \label{eq:St}
\end{gather}
where $\overline{F}_D$ is the mean drag force on the plate,
$A = s d$ is the flow-normal projected area of the plate  and $f$ is the vortex shedding frequency.  

Table~\ref{tab:constud} summarizes the grid refinement studies. First, we use the FVM with 2D simulations to assess if the discretization of the $x$--$y$ plane is sufficiently refined. 
Changes in both $\overline{C}_D$ and $St$ drop to below 2\% from case 2D\_4 to case 2D\_5. The mesh 2D\_5 is deemed sufficiently refined and is extruded along the spanwise dimension to produce 3D meshes referred to as 3D\_1--DNS400V in Table~\ref{tab:constud}. The spanwise refinement from case 3D\_5 to DNS400V results in a change of less than 1\% in both $\overline{C}_D$ and $St$. 

The base mesh for the SEM simulations is a coarsened version of the DNS400V mesh. The refinement, or increase in the number of degress of freedom, is achieved by increasing the polynomial order of the spectral elements ($p$-refinement).
It is expeditious to pair $p$-refinement with increasingly finer spatial discretization ($h$-refinement)~\cite{pozrikidis2014,karniadakis2005}.
The first $p$-refinement from P5\_1 to P7\_1 results in a 7.6\% change in $\overline{C}_D$ and $St$. Successive $h$ and $p$-refinements produce smaller changes
from P7\_2 to DNS400. 
Minor differences (in the third significant digit) in $\overline{C}_D$ and $St$ between the cases DNS400V and DNS400 are attributed to the different numerical methods employed.

Convergence of the numerical solutions with respect to the time discretization was verified by performing computations with different time steps $\Delta t$, all chosen to satisfy the CFL condition required for numerical stability. The time step $\Delta t$ was chosen to keep the CFL number around 0.6, corresponding to $\Delta t=0.001$ for DNS400V and $\Delta t = 0.002$ for DNS400.

\begin{table}
    \centering
    \begin{tabular}{ll|lllll}\hline\hline
               & Case     & $N_{xy}$ & $N_z$ & $p$-order & $\overline{C}_D$ & $St$ \\\hline
        2D FVM & 2D\_1    & 12576    & -     & -         & 2.402            & 0.174 \\
               & 2D\_2    & 18046    & -     & -         & 2.374            & 0.172 \\
               & 2D\_3    & 23048    & -     & -         & 2.327            & 0.164 \\
               & 2D\_4    & 33431    & -     & -         & 2.395            & 0.162 \\
               & 2D\_5    & 47000    & -     & -         & 2.418            & 0.160 \\
               & 2D\_6    & 60224    & -     & -         & 2.423            & 0.160 \\
               & 2D\_7    & 77987    & -     & -         & 2.425            & 0.160 \\
               & 2D\_8    & 101519   & -     & -         & 2.426            & 0.160 \\\hline
        3D FVM & 3D\_1    & 47000    & 67    & -         & 2.247            & 0.158 \\
               & 3D\_2    & 47000    & 80    & -         & 2.098            & 0.159 \\
               & 3D\_3    & 47000    & 100   & -         & 2.014            & 0.158 \\
               & 3D\_4    & 47000    & 134   & -         & 1.999            & 0.160 \\
               & 3D\_5    & 47000    & 200   & -         & 2.002            & 0.159 \\
               & DNS400V  & 47000    & 400   & -         & 1.984            & 0.158 \\\hline
        3D SEM & P5\_1    & 14472    & 48    & 5         & 1.567            & 0.177 \\
               & P7\_1    & 25728    & 64    & 7         & 1.452            & 0.164 \\
               & P5\_2    & 14472    & 240   & 5         & 1.935            & 0.150 \\
               & P7\_2    & 25728    & 320   & 7         & 1.912            & 0.149 \\
               & DNS400   & 40200    & 400   & 9         & 1.908            & 0.152 \\
        \hline\hline
    \end{tabular}
    \caption{Results of mesh convergence and $p$-refinement studies at $Re=400$.}
    \label{tab:constud}
\end{table}

\begin{figure}
    \centering
    \mbox{\subfigure[]{\includegraphics[scale=0.5]{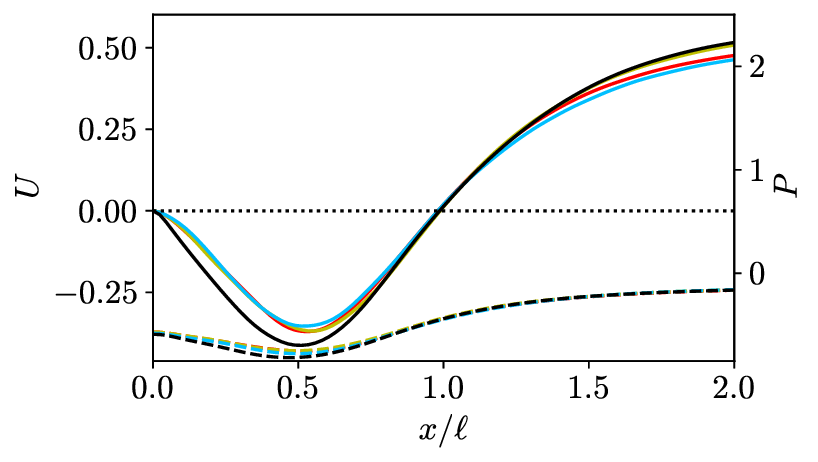}} 
    \subfigure[]{\includegraphics[scale=0.5]{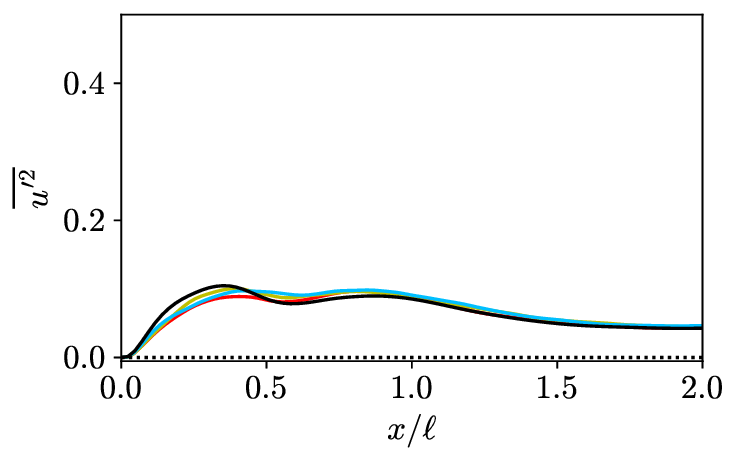}}}\\
    \mbox{\subfigure[]{\includegraphics[scale=0.5]{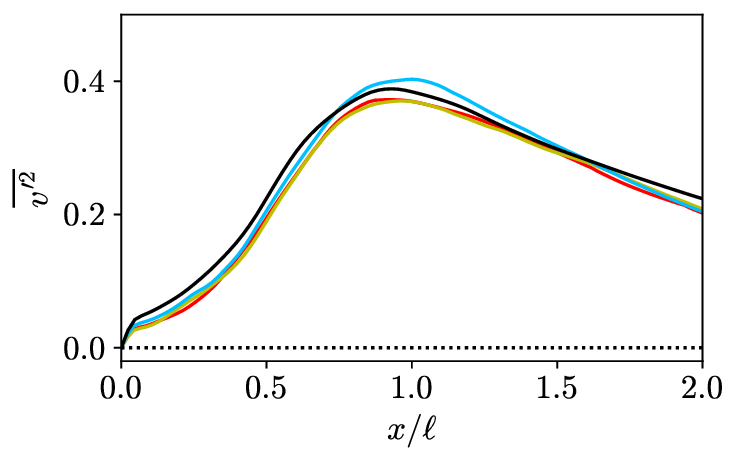}}
    \hspace{0.5cm}\subfigure[]{\includegraphics[scale=0.5]{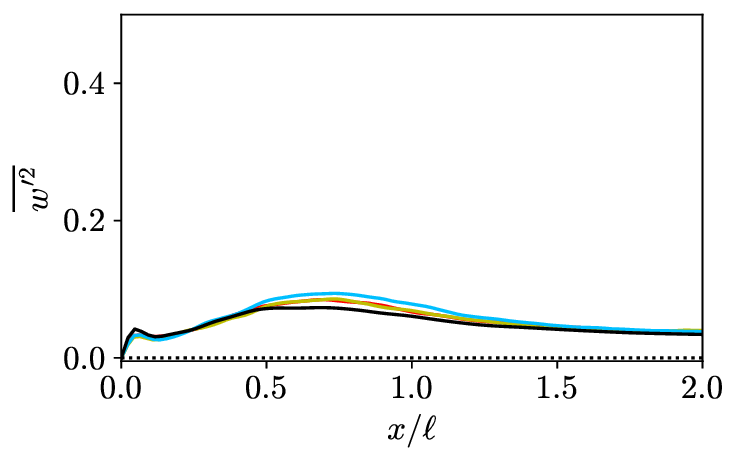}}}    
    \caption{Centreline profiles of the time-averaged quantities (a) $U$ (solid lines) and $P$ (dashed lines), (b) $\overline{u^{\prime 2}}$, (c) $\overline{v^{\prime 2}}$ and (d) $\overline{w^{\prime 2}}$ in the simulations performed on different meshes, cf.~Table \ref{tab:constud}: P5\_2 \Plot{red,solid}; P7\_2 \Plot{yellow,solid}; DNS400 \Plot{cyan,solid}; DNS400V \Plot{black,solid}.}
    \label{fig:centrevalidprofs}
\end{figure}

Finally, as is evident from Figures \ref{fig:centrevalidprofs}(a--d), the evolution of the mean streamwise velocity $U$, mean pressure $P$, and of the Reynolds stresses $\overline{u^{\prime 2}}$, $\overline{v^{\prime 2}}$ and $\overline{w^{\prime 2}}$ on the downstream of the plate along the flow centreline is very similar in the solutions obtained with the FVM and the SEM approaches on the most refined meshes and the same observation also applies to the dependence of these quantities on the transverse coordinate (not shown here for brevity).
This thus allows us to conclude that the simulations are well resolved, cf.~Table \ref{tab:constud}.

\subsection{Resolution of Kolmogorov Scale}\label{sub:kolm}

As a final test of the $Re=400$ CFD model's ability to accurately resolve turbulence, the Kolmogorov length scale $\eta$ is calculated at several points along the centreline, with
\begin{gather}
    \eta = \left(\frac{\nu^3}{\epsilon}\right)^{1/4},
\end{gather}
where $\epsilon$ is the homogeneous dissipation rate computed directly as
\begin{gather}
    \epsilon = \nu\overline{\frac{\partial u_i^\prime}{\partial x_j}\frac{\partial u_i^\prime}{\partial x_j}},\label{eq:2.12}
\end{gather}
or as the residual of the turbulent kinetic energy transport equation \eqref{eq:ktrans}.
In \eqref{eq:2.12}, $i,j=1,2,3$, $x_j$ represents the Cartesian coordinate (denoted $x$, $y$, or $z$ elsewhere), $u'_j$ represents the fluctuating velocity components (denoted $u'$, $v'$, or $w'$ elsewhere), and Einstein's convention was used with repeated indices implying summation. In this study, we use both methods and then compare the quantity $\Delta x/\eta$ with several other studies which use similar geometries and $Re$ in Table \ref{tab:kolmogorov}. Both our direct and residual calculations show similar Kolmogorov length scale resolutions in the near wake to those of the literature. Therefore, we consider the ability of DNS400 to accurately resolve turbulent phenomena validated.
\begin{table}[H]
    \centering
    \begin{tabular}{l|l|llll}\hline\hline
        Study                       & $Re$      & $x=$ & 1    & 2    & 5    \\\hline
        DNS400 (direct)            & 400       &      & 1.69 & 2.02 & 2.61 \\
        DNS400 (residual)          & 400       &      & 1.79 & 2.09 & 2.57 \\\hline
        Freeman et al. \cite{freeman2023}          & 250       &      & $<3.3$ & $<3.3$ & $<3.3$ \\
        Narasimhamurthy and Andersson \cite{narasimhamurthy2009}  & 750       &      & 3.48 & 2.69 & 2.19 \\
        Yao et al. \cite{yao2001}              & 1000      &      & 6.88 & 5.07 & 3.42 \\
        Hemmati et al. \cite{hemmati2016}          & 1200      &      & 1.85 & 2.59 & 3.24 \\\hline\hline
    \end{tabular}
    \caption{Comparison of $\Delta x / \eta$ for DNS400 simulation with other numerical studies on rectangular cylinders.}
    \label{tab:kolmogorov}
\end{table}

Following the convergence and validation studies of DNS400, 4096 snapshots of velocity and pressure fields are collected, each spaced 164 time steps apart. This encompasses 200 vortex shedding periods with approximately  20 snapshots per period. All flow statistics for DNS400 are computed using these 4096 snapshots. It was determined that 200 vortex shedding cycles were sufficient to obtain statistical stationarity.

DNS400 makes use of the modified polynomial shape functions of {\tt Nektar++}, as well as the spectral/$hp$ de-aliasing. 
The minimum number of quadrature points $Q_p$~\cite{karniadakis2005} necessary to de-alias (\ref{eq:mom}) and (\ref{eq:inc}) is
\begin{gather}
    Q_p=\frac{3p}{2} + 2
\end{gather}
where $p$ is the degree of the approximating polynomials.
DNS400 makes use of polynomial solutions of degree 9, therefore, $Q_p=16$.

\subsection{Experimental Data}

In this study we use two experimental data sets, both corresponding to a fully turbulent regime, for comparison with the DNS computations described above.
For both data sets, the three velocity components were measured simultaneously along a $xy$-plane in the plate near wake using time-resolved stereoscopic particle image velocimetry (PIV). The experiments were conducted in the high-contraction ratio, open jet suction-type wind tunnel of the Laboratory for Turbulence Research in Aerodynamics and flow Control (LTRAC) at the University of Calgary. The laboratory ambient is controlled and all experiments were conducted at the nominal atmospheric conditions of $20^\circ$C and 90~kPa. The corresponding fluid properties are $\nu=1.71\times10^{-5}$~$\mbox{m}^2$/s and $\rho=1.05$~kg/$\mbox{m}^3$. The test facility and PIV system are described in detail in \cite{braun2020} and only key features are briefly summarized here.

The two experiments differed in the plate dimensions and test conditions.
The first one, referred to as EX12K, is from Rashid \cite{ashkerphd}.
The plate dimensions were: $d=0.0128$~m; the wetted span $s=0.457$~m; and thickness $b=6.67\times 10^{-4}$~m. The experiments were conducted for a nominal $U_\infty=16.8$m/s and $Re=12500$. The second, EX20K, is from~\cite{braun2020}, corresponds to a nominal Reynolds number of 20000 with parameters: 
$d=0.0224$~m; $s=0.508$~m; $b=4.40\times 10^{-3}$~m; $U_\infty=14.8$m/s for $Re=19700$.
Both test-plates consisted of machine-flat steel. Square rigid end-plates were mounted at $\pm s/2$ of the test plate. These end-plates were nominally parallel to the free-stream and are adjusted (within $\pm 2^\circ$) to ensure 2D mean flow conditions in the wake. The frontal aspect ratio, $s/d$ for EX12K and EX20K is 36 and 23, respectively, thus allowing to verify that the influence of $s/d$ can be neglected.

Instantaneous three-component velocity fields were collected in the midplane of the plate utilizing a \textit{Lavision} PIV system. The flow was illuminated using a \textit{Photonics Industries} 20~mJ Nd:YLF (532~nm) laser and image pairs were captured with {Photron} SA4 Fastcams). The cameras were mounted at $45^\circ$ relative to the measurement plane. The flow was seeded with olive oil particles, of nominal mean diameter 1~$\mu$m, introduced upstream of the wind tunnel inlet. The nominal PIV field of view is approximately $5.5d\times5.5d$. Image pairs (snapshots) were sampled between 1~KHz and 2~KHz, or about 12 to 15 snapshots per shedding cycle. A measurement trial consisted of 2728 snapshots, or about 180 shedding cycles. Each measurement consisted of six independent trials. PIV images were processed using \textit{Davis Flow Master 8.10} software. The mean velocity and Reynolds stress fields are the ensemble averages of the six trials.
The estimated measurement uncertainty for $U$ and $V$ is $\pm 0.01$; for $W$, $\pm 0.03$; for $\overline{u'^2}, \overline{v'^2}$ and $\overline{u'v'}$, $\pm 0.02$ and for $\overline{w'^2}$, $\pm 0.04$ 
Table \ref{tab:cases} summarizes key information about the DNS and experimental data sets used in this study.

\begin{table}[H]
    \centering
    \begin{tabular}{l|ll}\hline\hline
         Name & $Re$ & Description \\\hline
         DNS150 & 150 & 3D SEM DNS\\
         DNS400V & 400 & 3D FVM DNS used for mesh convergence and validation\\
         DNS400 & 400 & 3D SEM DNS\\
         EX12K  & 12500 & Stereo PIV wind tunnel experiment\\
         EX20K  & 19700 & Stereo PIV wind tunnel experiment\\\hline\hline
    \end{tabular}
    \caption{Summary of the DNS and experimental data sets used in the present study.}
    \label{tab:cases}
\end{table}



\subsection{Spectral Analysis}

Frequency energy spectra of the fluctuating velocity components $(u',v',w')$ are estimated as the Fourier coefficients, $(\hat{u},\hat{v},\hat{w})$, obtained from a discrete Fourier transform (DFT) using an FFT algorithm with $N=4096$ points. where the time series is sampled at $f_s \approx St/20.48$. These are calculated using the method of Welch~\cite{welch1967} with a window weighting function $w_i$  and averaged over ten overlapping time segments. The power spectral density function, for example for $u'$, is then defined as:
\[
\Phi_{uu}(f_i) = \frac{|\hat{u_i}|^2}{f_s(\frac{1}{N} \Sigma_{i=1}^{N-1} w^2_i)}, \, \hspace{1cm} 
w_i = 1- \big[ (2\times i /(N-2) - 1 \big]^2 , 
\]
thus satisfying Parseval's identity $\overline{u'^2} = \int_{-\infty}^{\infty} \Phi_uu(f) \mbox{d}f \approx \frac{1}{N} \Sigma_{i=0}^{N-1} \Phi_{uu}(f_i)$.

\FloatBarrier
\section{Results}\label{sec:results}

The purpose of this section is to provide evidence that the wake flow past a thin flat plate is in a turbulent state already at $Re = 400$ and that changes with icreasing $Re$ are slow. 
This is done by performing careful quantitative comparisons of the DNS of this flow (DNS400) with the DNS of the flow at a lower Reynolds number (DNS150) and  experimental studies of two wake flows at higher Reynolds numbers (EX12K and EX20K), all for the same nominal geometry, cf.~Table \ref{tab:cases}. Section \ref{sub:globQs} analyzes the global (integral) quantities associated with these wakes. The corresponding mean velocity and the Reynolds stress fields are discussed in Section \ref{sub:meanvels}. The local energy budgets are studied in Section~\ref{sub:budgets} and energy spectra of the fluctuating velocity fields in Section \ref{sub:spects}. In Section \ref{sub:hist}, we discuss the statistical properties of the symmetric and antisymmetric parts of the velocity gradient tensor $\nabla\mathbf{u}$ and, finally, the spatial properties of the far wakes are analyzed in Section \ref{sub:farwake}. 

\subsection{Global Variables}\label{sub:globQs}

We focus on the mean drag coefficient  $\overline{C}_D$, the nondimensional shedding frequency $St$ and the mean recirculation length $\ell$ (nondimensionalized by $d$) representing the downstream distance from the back of the plate to the point on the flow center plane where the mean streamwise velocity $U$ changes sign. These quantities are reported in Table \ref{tab:globQs} for the different flows considered here, as well as for several studies from the literature spanning a wide range of Reynolds numbers.
At first glance, the quantities $\overline{C}_D$ and $St$, reveal no clear dependence on $Re$. For $\overline{C}_D$, the studies above the 3D transition report values in the range 1.9-2.5. For $St$, the values range from 0.14 to 0.17. On the other hand, $\ell$ does show a clearer trend with values beginning at around 2 just above the 3D transition and generally increasing with $Re$, albeit slowly. 
 
Most CFD studies were conducted for periodic spanwise boundary conditions such that the two-dimensionality of the mean flow was enforced by construction.
For experimental studies, the spanwise boundary condition consists typically of end plates. This boundary condition was enforced in some CFD studies~\cite{hemmati2016,freeman2023}.
For both CFD cases and the experiments EX20K~\cite{braun2020} and EX12K~\cite{ashkerphd}, it was confirmed that the mean spanwise gradients of the velocity and higher statistical moments are negligible over 80\% of the span about $z=0$.  The present DNS at $Re=400$, that of Joshi~\cite{joshi1993} at $Re = 1000$ and the LES study of Tian et al.~\cite{tian2014} at $Re=150000$ give values of the integral parameters closest to those observed experimentally. The comparison of this subgroup of studies suggests that for $Re>400$, $\overline{C_D}, St$ and $\ell$ are slow varying functions of $Re$.

\begin{table}[H]
    \centering
    \begin{tabular}{l|l|l|lll}\hline\hline
        \rule{0pt}{2.5ex}Case & Type & $Re$    & $\overline{C}_D$ &  $St$    & $\ell$ \\
        \hline        
        DNS150 & DNS (3D) & 150 & 2.29 & 0.144 & 1.594 \\
        DNS400 & DNS (3D) & 400     & 1.91            & 0.152   & 2.273 \\
        EX12K \cite{ashkerphd} & EX & 12458   & -                & 0.145   & 2.16 \\
        EX20K \cite{braun2020} & EX & 20000   & 2.10                & 0.142   & 2.44 \\\hline
        Najjar \& Balachandar \cite{najjarbalachandar1998} & DNS (3D) & 250 & 2.36 & 0.161 & 2.350 \\
        Singh and Narasimhamurthy \cite{singh2022} & DNS (3D) & 250 & 2.22 & 0.164 & 2.040 \\
        Freeman et al. \cite{freeman2023} & DNS (3D) & 250 & 2.29 & 0.165 & 2.06 \\
        Ong \& Yin \cite{ongyin2022} & DNS (3D) & 500, 1000 & 2.07 & 0.16 & - \\
        Afgan et al. \cite{afgan2013} & LES (3D) & 750 & 2.29 & 0.167 & 1.880 \\
        Narasimhamurthy and Andersson \cite{narasimhamurthy2009} & DNS (3D) & 750 & 2.31 & 0.168 & 1.96 \\
        Joshi \cite{joshi1993} & DNS (3D) & 1000 & 2.47 & 0.150  & 2.300 \\
        Hemmati et al. \cite{hemmati2018} & DNS (3D) & 1200 & 2.13 & 0.158 & 2.65 \\
        Fage and Johansen \cite{fage1927} & EX & 150000 & 2.13 & 0.146 & - \\
        Tian et al. \cite{tian2014} & LES (3D) & 150000 & 2.31 & 0.155 & 2.25 \\\hline\hline
    \end{tabular}
    \caption{Comparison of the global quantities the mean drag coefficient $\overline{C}_D$, Strouhal number $St$, and recirculation length $\ell$) characterizing the thin flat plate wakes considered in Table \ref{tab:cases} and literature.}
    \label{tab:globQs}
\end{table}

\subsection{Mean Velocity and Reynolds Stress Fields}\label{sub:meanvels}

\begin{figure}
    \centering
    \mbox{
        \subfigure[]{\includegraphics[scale=0.5]{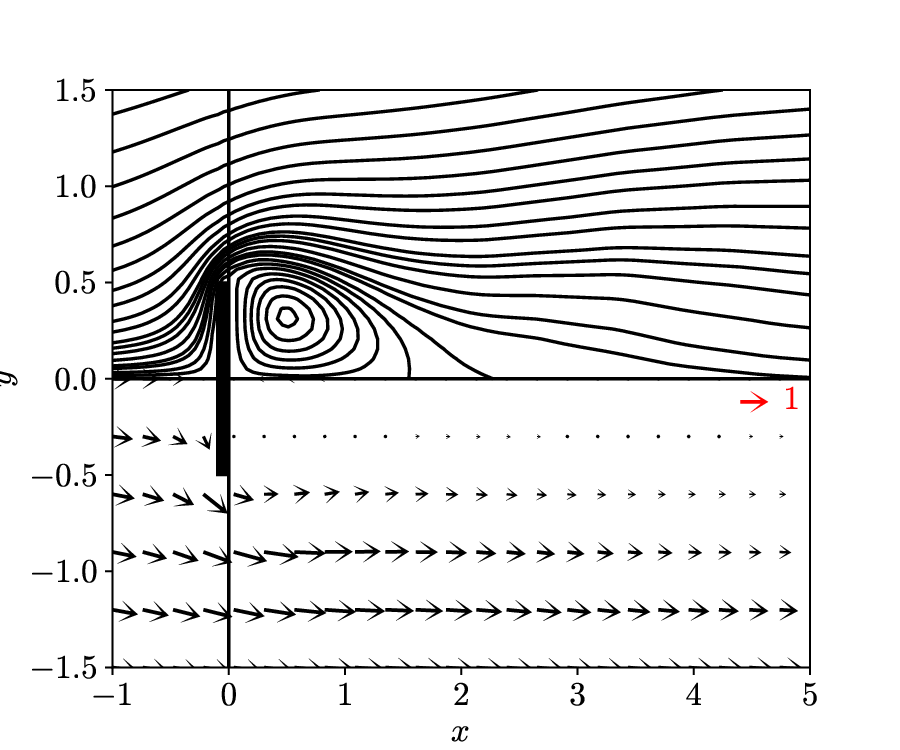}}\label{fig:streamvecs150}
        \subfigure[]{\includegraphics[scale=0.5]{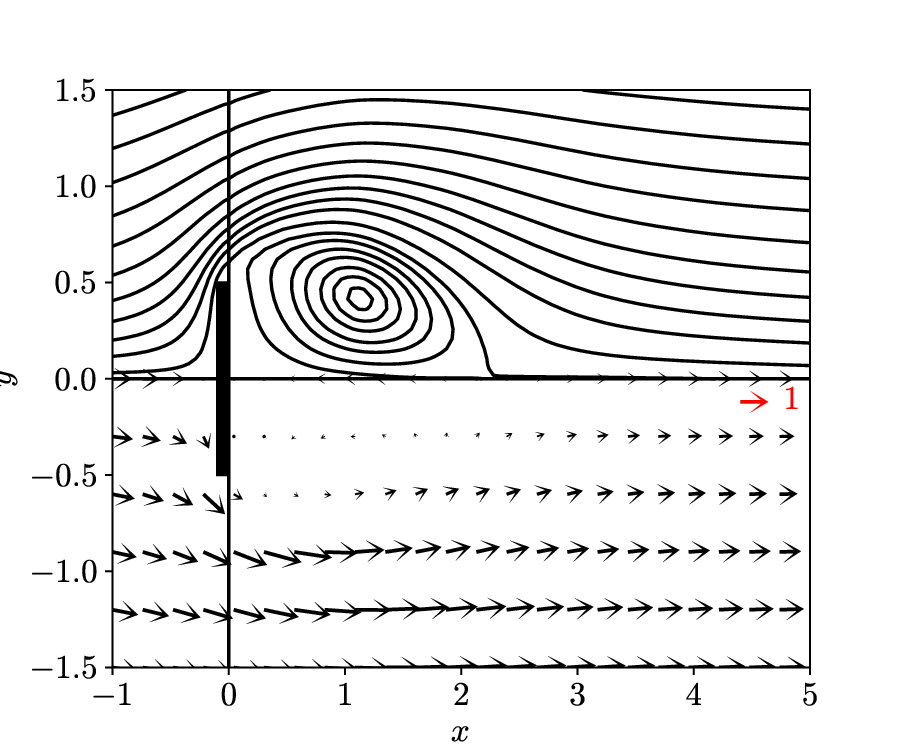}}\label{fig:streamvecs400}}
    \mbox{
        \subfigure[]{\includegraphics[scale=0.5]{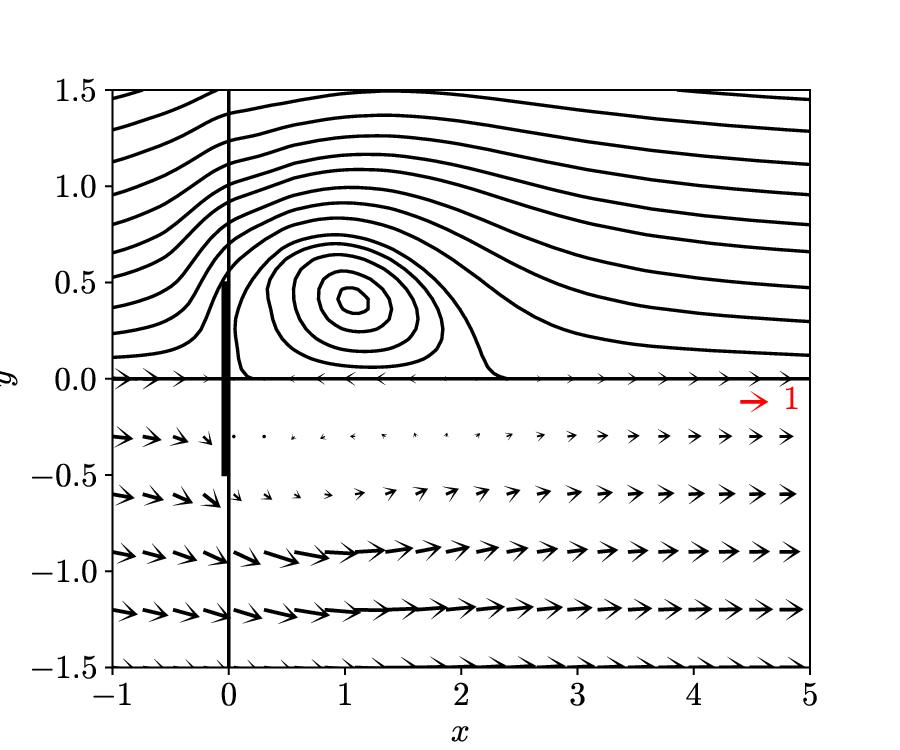}}\label{fig:streamvecs12458}
        \subfigure[]{\includegraphics[scale=0.5]{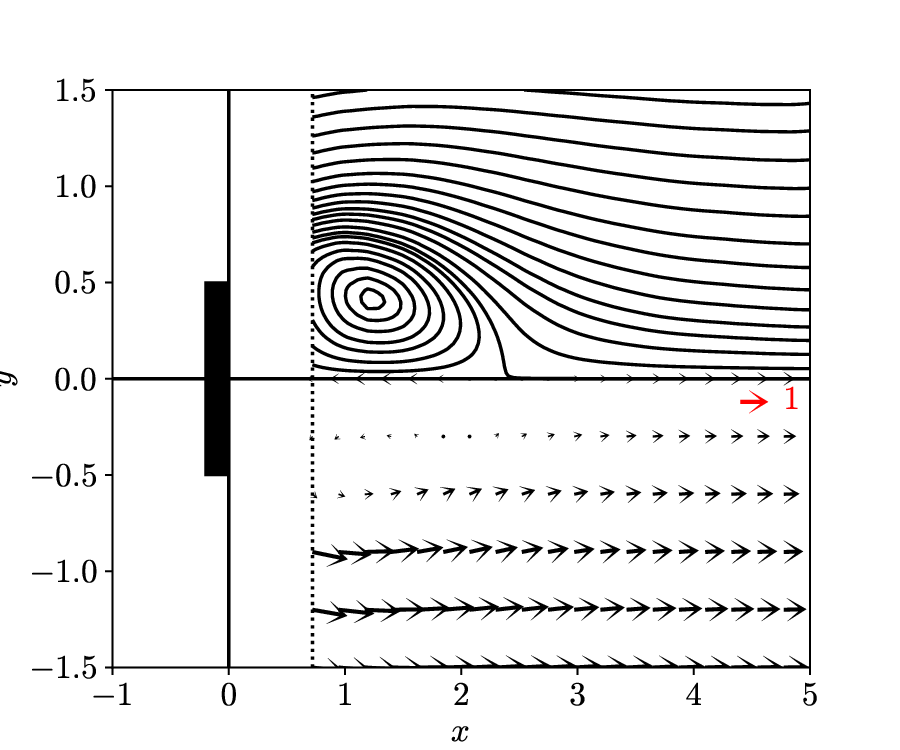}}\label{fig:streamvecs20000}}
\caption{Mean streamlines (top half planes) and velocity vector fields (bottom half planes) on the symmetry plane $y=0$  for (a) DNS150, (b) DNS400, (c) EX12K and (d) EX20K. Flat plates are depicted in all figures with streamwise thicknesses to scale.
Streamlines are iso-contours of the stream function $\Psi = \int_0^y u \mbox{d}y$.}\label{fig:streamvecs}
\end{figure}

\begin{figure}
    \centering
    \includegraphics[scale=0.5]{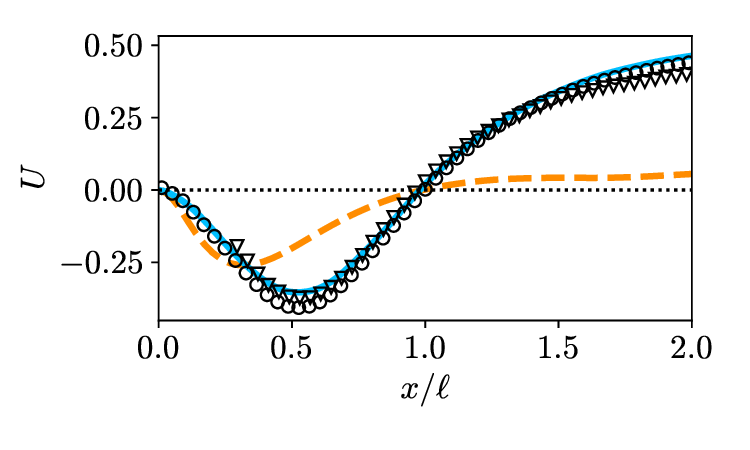}
    \caption{Centreline profiles of the mean streamwise velocity $U(x,0,0)$. Legend: DNS150 \Plot{orange,dashed}; DNS400 \Plot{cyan,solid}; EX12K ($\bigcirc$); EX20K ($\bigtriangledown$).}
    \label{fig:u_centreprofs}
\end{figure}
Mean streamlines and velocity vector fields in the flow center plane $z=0$ are shown in Figure \ref{fig:streamvecs} for DNS150, DNS400, EX12K and EX20K. 
Since both fields are symmetric with respect to the axis $y = 0$, they are shown in half-planes only. We note that the spanwise direction is homogeneous such that the mean spanwise velocity component identically vanishes in the flow center plane, $W \equiv 0$, and the mean flow is two-dimensional. For $Re \ge 400$, the streamline topology is similar in all cases, albeit predictions of the location of critical points show some small scatter; for example, the saddle point indicating the extent of the mean recirculation region $\ell$ as indicated in Table \ref{tab:globQs}.
In contrast, for $Re=150$ the recirculation length is markedly shorter. The streamlines show inflections in the near wake and tend to diverge for $|y|>1$. These structural differences hint at differences in dynamics for this case compared to $Re \ge 400$.

The variation of the mean streamwise velocity $U$ along the flow axis in the different cases is shown as function of $x/\ell$ in Figure \ref{fig:u_centreprofs}. In all cases, the location $x$ at which $U$ is minimum for $x< \ell$ corresponds to that of the recirculation nodes (centers) in Figure~\ref{fig:streamvecs}. The variation is similar in the cases DNS400, EX12K and EX20K. These feature a strong recirculation region followed by a wake recovery. The recirculation nodes are located at $x/\ell \approx 0.5$ and the inflection point, where $\partial U/\partial x$ is maximum, at $x\approx \ell$. On the other hand, for $Re=150$ the structure of the flow field differs. The backflow along the axis $y = 0$  is much weaker and the wake recovery much slower. The recirculation nodes are much closer to the plate, $x/\ell \approx 0.3$, and the inflection point occurs within the recirculation, $x/\ell \approx 0.7$.

\begin{figure}[H]
    \centering
    \mbox{
        \subfigure[]
        {\includegraphics[scale=0.5]{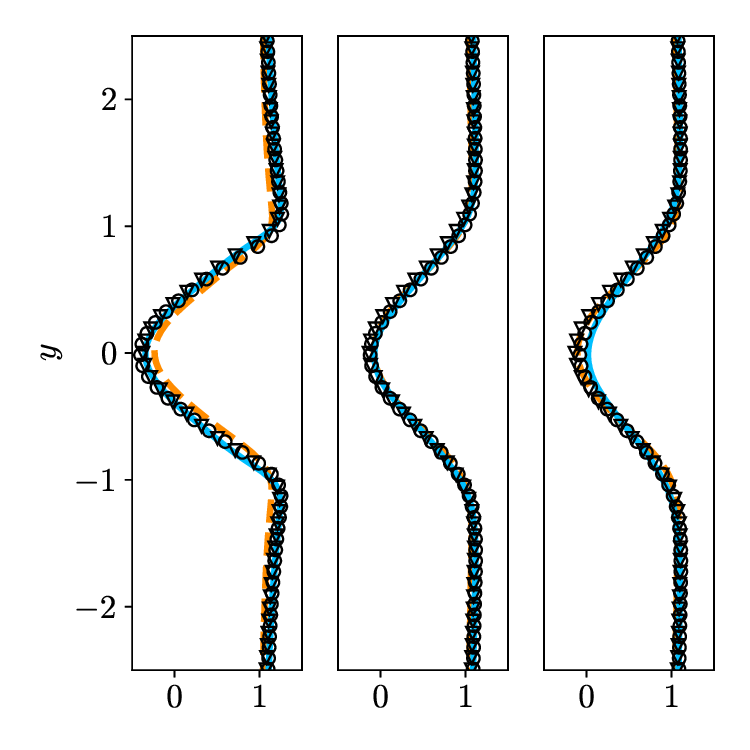}}
        \subfigure[]
        {\includegraphics[scale=0.5]{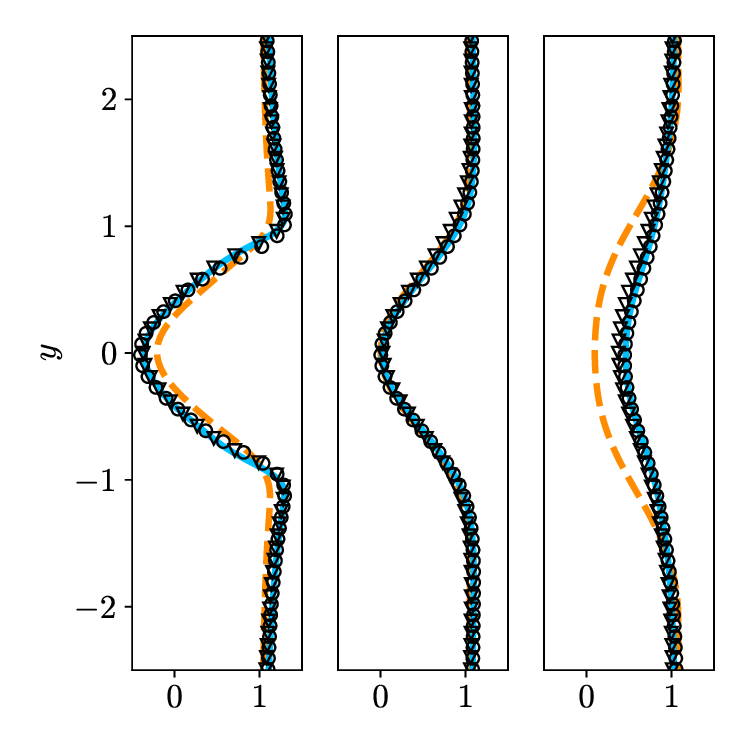}}}
\caption{Transverse profiles of the mean streamwise velocity $U(x_p,y,0)$ (a) at the downstream locations $x_p$ where the maxima of the Reynolds stresses (from left to right) $\overline{u^{\prime 2}}$, $\overline{u^\prime v^\prime}$ and $\overline{v^{\prime 2}}$ are attained, and (b) at (from left to right) $x_p = \ell/2$, $x_p = \ell$ and $x_p = 4.5$, cf.~Table \ref{tab:proflocs}. Legend: DNS150 \Plot{orange,dashed}; DNS400 \Plot{cyan,solid}; EX12K ($\bigcirc$); EX20K ($\bigtriangledown$).}
\label{fig:u_vertprofs}
\end{figure}

\begin{figure}
    \centering
    \mbox{
        \subfigure[]
        {\includegraphics[scale=0.5]{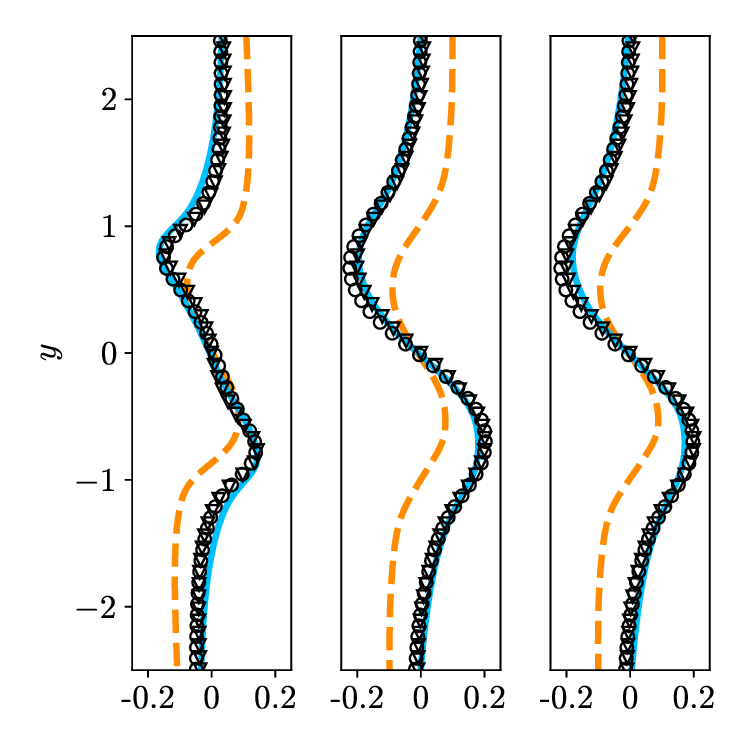}}
        \subfigure[]
        {\includegraphics[scale=0.5]{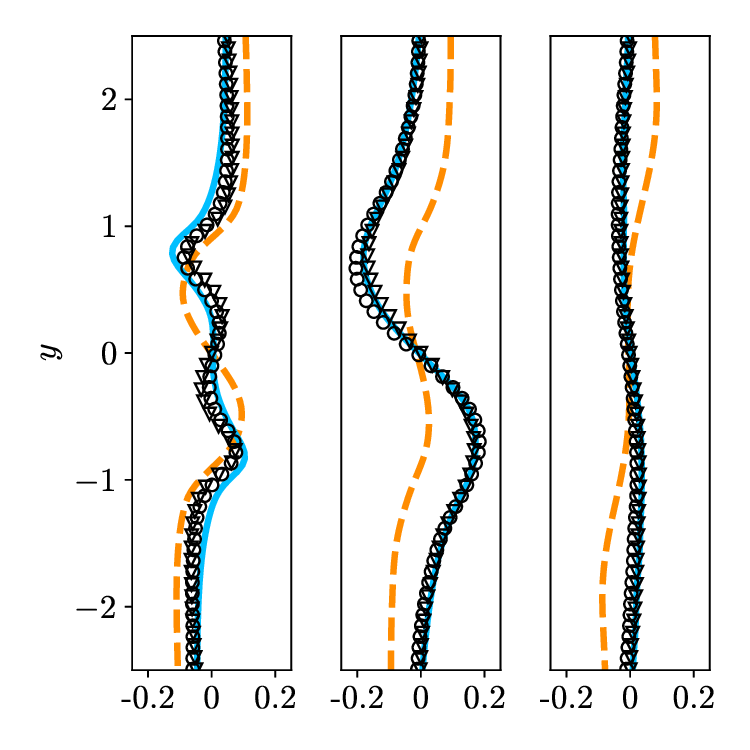}}}
\caption{Transverse profiles of the mean transverse velocity $V(x_p,y,0)$ (a) at the downstream locations $x_p$ where the maxima of the Reynolds stresses (from left to right) $\overline{u^{\prime 2}}$, $\overline{u^\prime v^\prime}$ and $\overline{v^{\prime 2}}$ are attained, and (b) at (from left to right) $x_p = \ell/2$, $x_p = \ell$ and $x_p = 4.5$, cf.~Table \ref{tab:proflocs}. Legend: DNS150 \Plot{orange,dashed}; DNS400 \Plot{cyan,solid}; EX12K ($\bigcirc$); EX20K ($\bigtriangledown$).}
\label{fig:v_vertprofs}
\end{figure}

Distributions of $U$ and $V$ along the transverse $y$ direction at different representative downstream locations specified in Table \ref{tab:proflocs} are shown in Figures \ref{fig:u_vertprofs} and \ref{fig:v_vertprofs}, respectively. As is evident from these plots, the behaviour of both mean velocity components is similar in the cases DNS400, EX12K and EX20K, and differs from that for DNS150. The differences in the $V$-distributions and in the downstream momentum deficit ($U$ component) indicate changes in transverse transport.

\begin{table}[H]
    \centering
    \begin{tabular}{r|llll}\hline\hline
        \rule{0pt}{2.5ex}Streamwise and transverse & DNS150 & DNS400 & EX12K & EX20K \\ locations $x_p$, $y_p$ &&&& \\
        \hline
        $\operatorname{argmax}_{x,y}\overline{u^{\prime 2}}(x,y,0)$ & (0.653,-0.732) & (1.364,-0.581) & (1.220,-0.527) & (1.392,-0.480) \\
        $\operatorname{argmax}_x\overline{v^{\prime 2}}(x,y,0)$ & (1.053,-0.682) & (2.273,-0.025) & (1.989,-0.015) & (2.064,0) \\
        $\operatorname{argmax}_x\overline{u^\prime v^\prime}(x,y,0)$ & (0.302,-0.682) & (2.071,-0.682) & (1.904,-0.613) & (2.064,-0.576) \\
        $\ell / 2$ & (0.797,0) & (1.137,0) & (1.080,0) & (1.224,0) \\
        $\ell$ & (1.594,0) & (2.273,0) & (2.160,0) & (2.448,0) \\
        4.5 & (4.5,0) & (4.5,0) & (4.5,0) & (4.5,0) \\\hline\hline
    \end{tabular}
    \caption{Locations within the plane $z=0$ where $y$-profiles of mean velocity components, Reynolds stresses, and terms of the $k$-transport equation are presented. The location of the different points $\mathbf{x}_p = [x_p, y_p, 0]$ in the flow DNS400 are illustrated in Figure \ref{fig:probelocs}.}
    \label{tab:proflocs}
\end{table}
\begin{figure}
    \centering
    \includegraphics[scale=0.5]{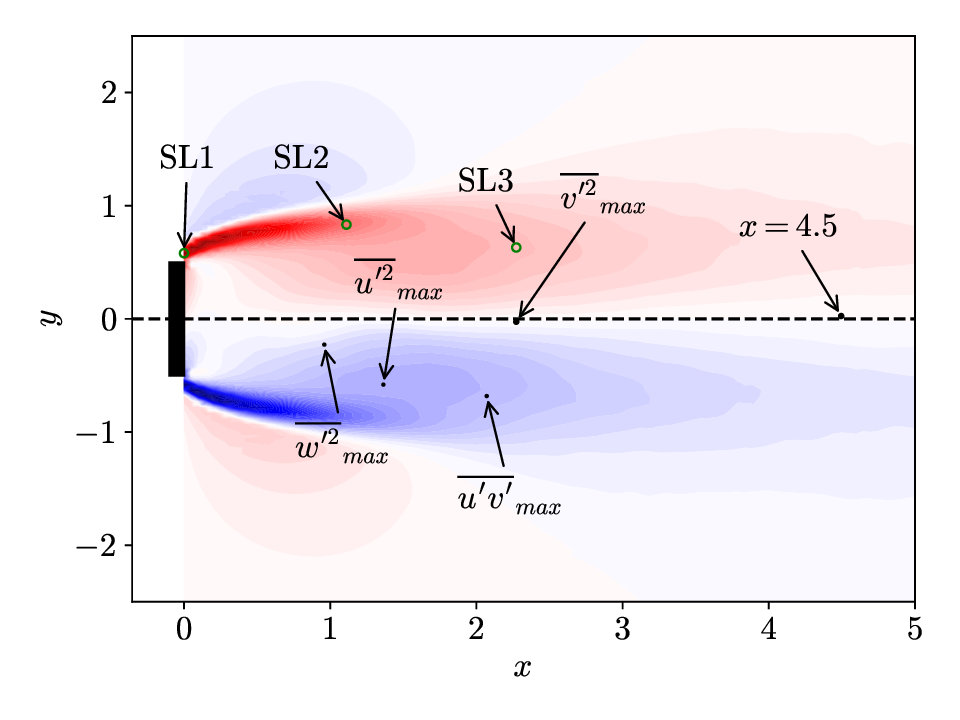}
    \caption{Locations $\mathbf{x}_p = [x_p, y_p,0]$ of the maxima  of the different Reynolds stresses and shear layer probes in the flow DNS400, cf.~Table \ref{tab:proflocs}. Green circles indicate probe locations in the shear layer (SL1--SL3), cf.~Table \ref{tab:probelocs2}, and black dots represent other locations of interest.}
    \label{fig:probelocs}
\end{figure}

\FloatBarrier

\begin{figure}
    \centering
    \subfigure[]{
    \includegraphics[scale=0.5]{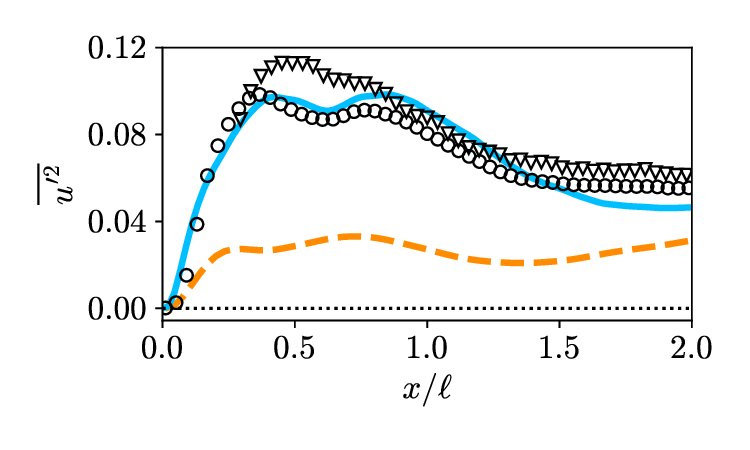}
    \label{fig:uu_centreprofs}}
    \subfigure[]{  
    \includegraphics[scale=0.5]{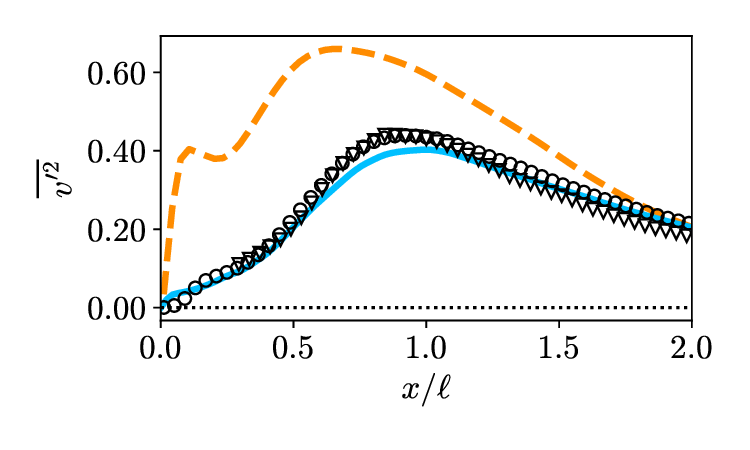}
    \label{fig:vv_centreprofs}}
    \subfigure[]{
    \includegraphics[scale=0.5]{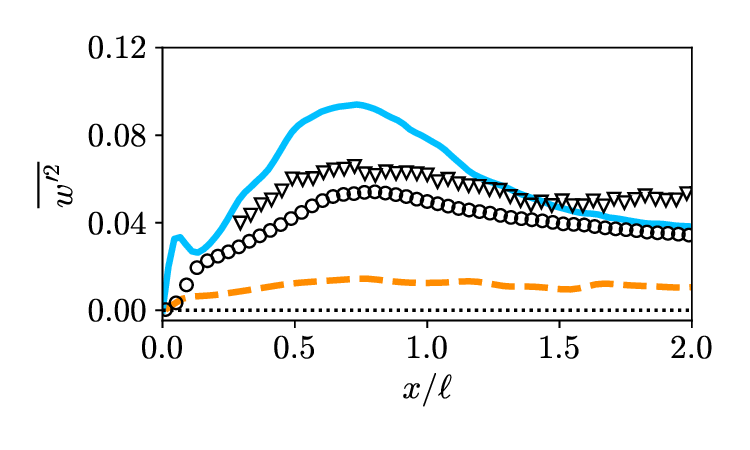}
    \label{fig:ww_centreprofs}}
    \caption{Centreline profiles of the Reynolds stresses (a) $\overline{u^{\prime 2}}(x,0,0)$, (b) $\overline{v^{\prime 2}}(x,0,0)$ and (c) $\overline{w^{\prime 2}}(x,0,0)$. Legend: DNS150 \Plot{orange,dashed}; DNS400 \Plot{cyan,solid}; EX12K ($\bigcirc$); EX20K ($\bigtriangledown$).}
    \label{fig:ReSt_centreprofs}
\end{figure}

We now examine the behaviour of the second-order moments characterizing turbulent flows, namely the Reynolds stresses.
While not shown for brevity, we have confirmed that $\overline{u'w'}$, $\overline{v'w'}$ and the $z$-gradients of the remaining Reynolds stresses vanish, within experimental uncertainty, for all cases as expected for 2D mean flows. 
The variation of $\overline{u^{\prime 2}}$, $\overline{v^{\prime 2}}$ and $\overline{w^{\prime 2}}$ with the downstream distance along the flow centre line is documented in Figure \ref{fig:ReSt_centreprofs}, noting that $\overline{u'v'}=0$ by symmetry. Whereas the longitudinal distributions of these Reynolds stresses is similar, generally within experimental uncertainty, for DNS400, EX12K and EX20K, these differ from those at $Re=150$ (DNS150). 

\begin{figure}
    \centering
    \mbox{
        \subfigure[]
        {\includegraphics[scale=0.5]{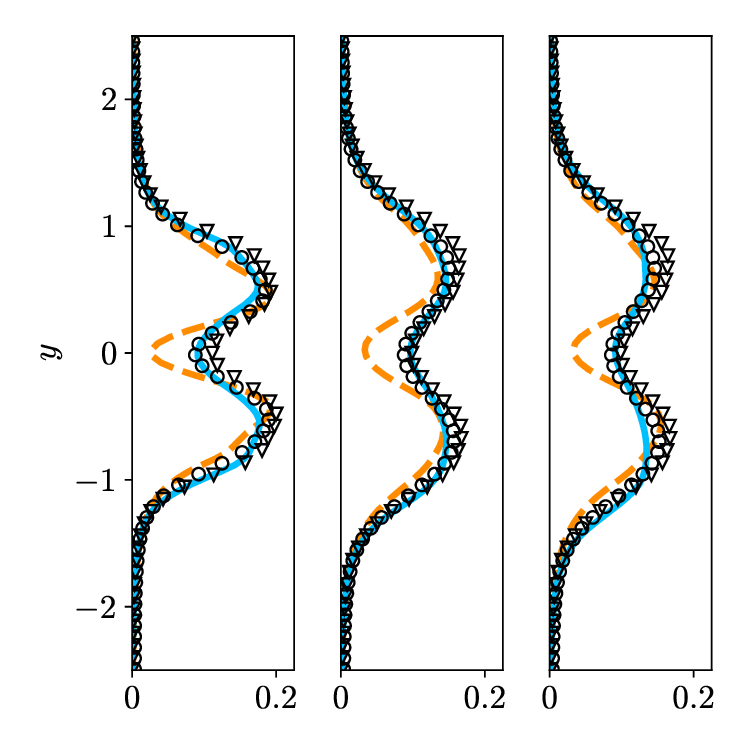}}
        \subfigure[]
        {\includegraphics[scale=0.5]{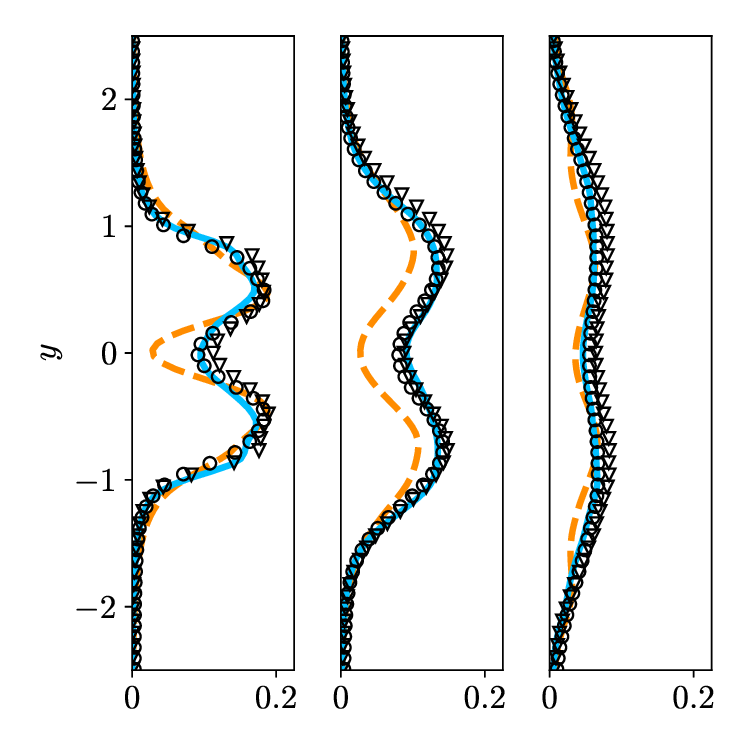}}}
\caption{Transverse profiles of the Reynolds stress $\overline{u^{\prime 2}}(x_p,y,0)$ (a) at the downstream locations $x_p$ where the maxima of the Reynolds stresses (from left to right) $\overline{u^{\prime 2}}$, $\overline{u^\prime v^\prime}$ and $\overline{v^{\prime 2}}$ are attained, and (b) at (from left to right) $x_p = \ell/2$, $x_p = \ell$ and $x_p = 4.5$, cf.~Table \ref{tab:proflocs}.  Legend: DNS150 \Plot{orange,dashed}; DNS400 \Plot{cyan,solid}; EX12K ($\bigcirc$); EX20K ($\bigtriangledown$).}
\label{fig:uu_vertprofs}
\end{figure}

\begin{figure}
    \centering
    \mbox{
        \subfigure[]
        {\includegraphics[scale=0.5]{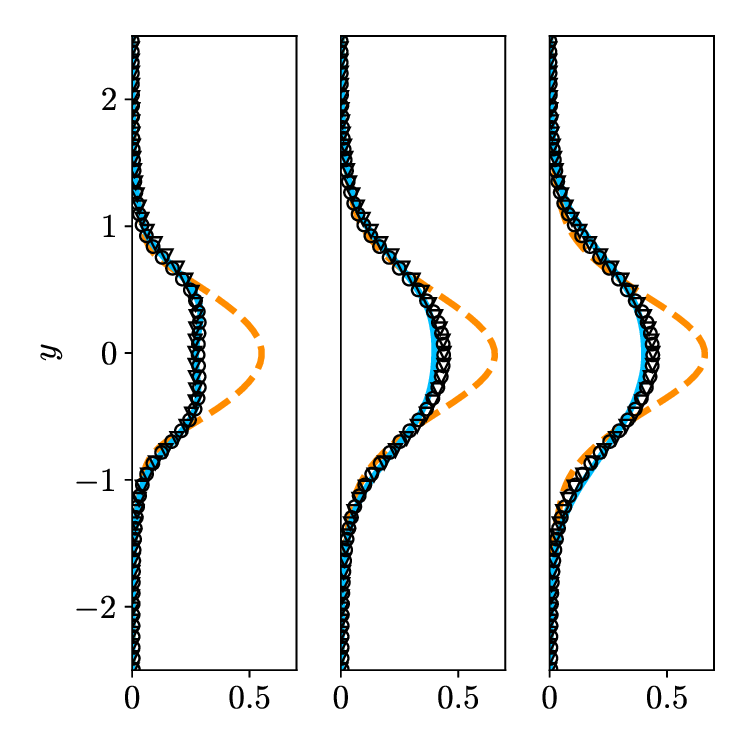}}
        \subfigure[]
        {\includegraphics[scale=0.5]{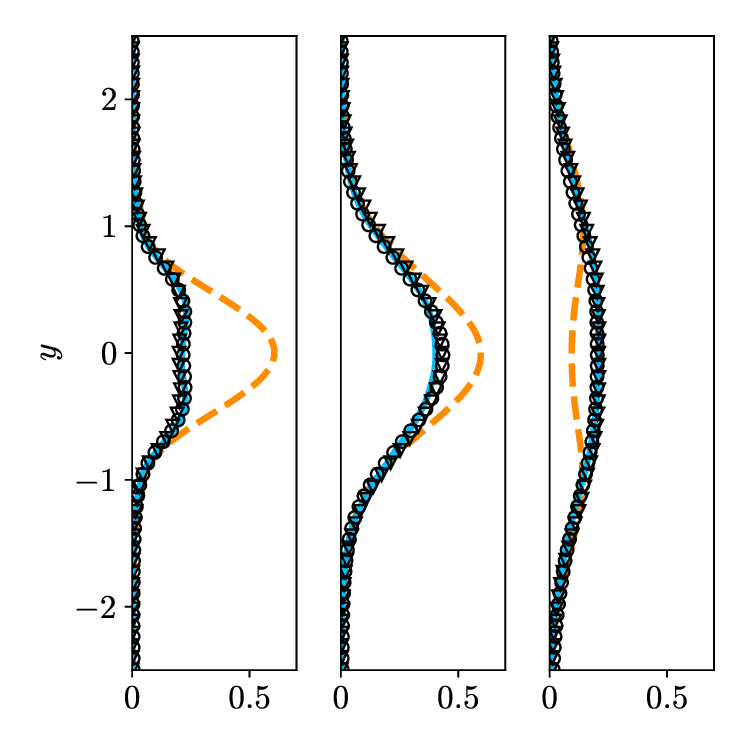}}}
\caption{Transverse profiles of the Reynolds stress $\overline{v^{\prime 2}}(x_p,y,0)$ (a) at the downstream locations $x_p$ where the maxima of the Reynolds stresses (from left to right) $\overline{u^{\prime 2}}$, $\overline{u^\prime v^\prime}$ and $\overline{v^{\prime 2}}$ are attained, and (b) at (from left to right) $x_p = \ell/2$, $x_p = \ell$ and $x_p = 4.5$, cf.~Table \ref{tab:proflocs}. Legend: DNS150 \Plot{orange,dashed}; DNS400 \Plot{cyan,solid}; EX12K ($\bigcirc$); EX20K ($\bigtriangledown$).}
\label{fig:vv_vertprofs}
\end{figure}

\begin{figure}
    \centering
    \mbox{
        \subfigure[]
        {\includegraphics[scale=0.5]{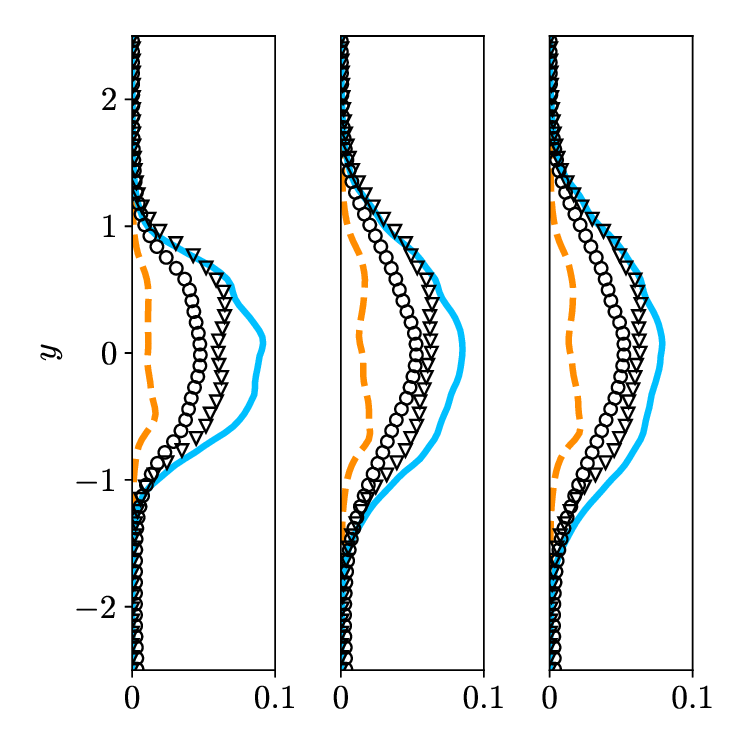}}
        \subfigure[]
        {\includegraphics[scale=0.5]{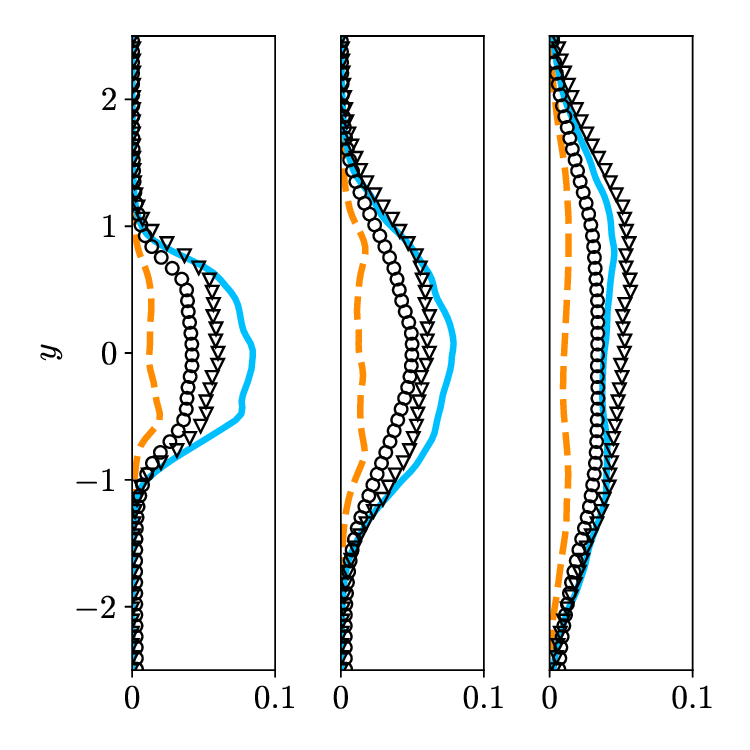}}}
\caption{Transverse profiles of the Reynolds stress $\overline{w^{\prime 2}}(x_p,y,0)$ (a) at the downstream locations $x_p$ where the maxima of the Reynolds stresses (from left to right) $\overline{u^{\prime 2}}$, $\overline{u^\prime v^\prime}$ and $\overline{v^{\prime 2}}$ are attained, and (b) at $x_p = \ell/2$, $x_p = \ell$, and $x_p = 4.5$ (from left to right), cf.~Table \ref{tab:proflocs}. Legend: DNS150 \Plot{orange,dashed}; DNS400 \Plot{cyan,solid}; EX12K ($\bigcirc$); EX20K ($\bigtriangledown$).}
\label{fig:ww_vertprofs}
\end{figure}

\begin{figure}
    \centering
    \mbox{
        \subfigure[]
        {\includegraphics[scale=0.5]{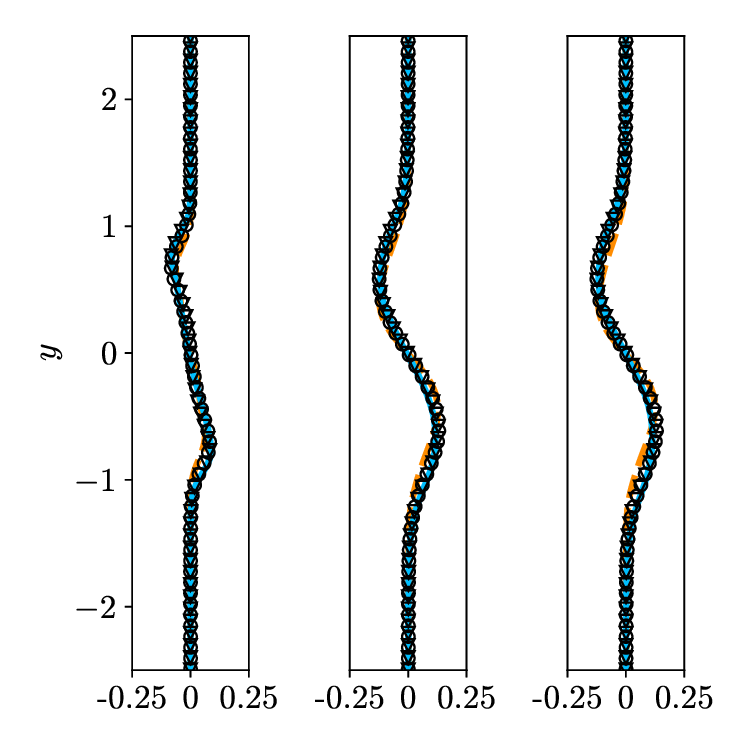}}
        \subfigure[]
        {\includegraphics[scale=0.5]{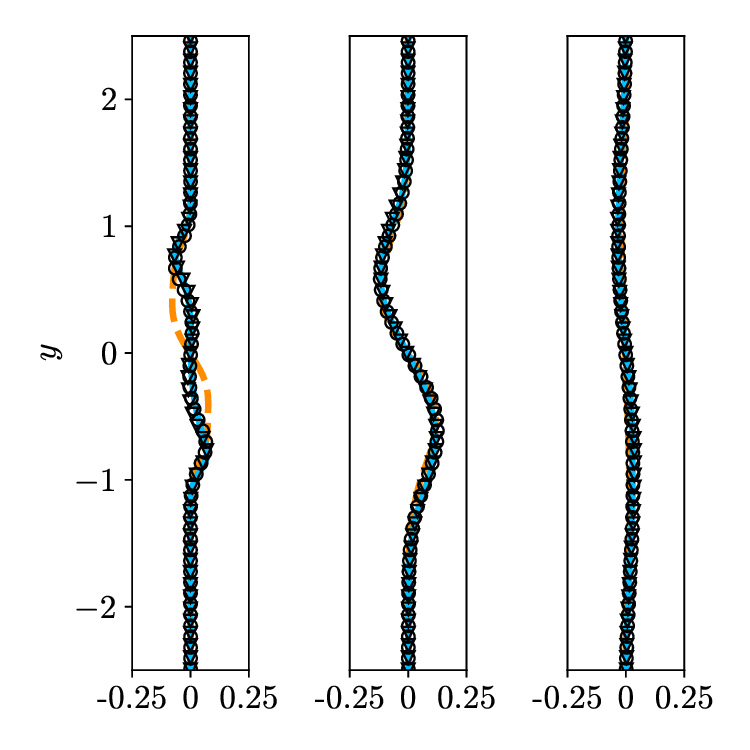}}}
\caption{Transverse profiles of the Reynolds stress $\overline{u^{\prime} v^{\prime}}(x_p,y,0)$ (a) at the downstream locations $x_p$ where the maxima of the Reynolds stresses (from left to right) $\overline{u^{\prime 2}}$, $\overline{u^\prime v^\prime}$ and $\overline{v^{\prime 2}}$ are attained, and (b) at (from left to right) $x_p = \ell/2$, $x_p = \ell$ and $x_p = 4.5$, cf.~Table \ref{tab:proflocs}. Legend: DNS150 \Plot{orange,dashed}; DNS400 \Plot{cyan,solid}; EX12K ($\bigcirc$); EX20K ($\bigtriangledown$).}
\label{fig:uv_vertprofs}
\end{figure}

The transverse profiles of the Reynolds stresses $\overline{u^{\prime 2}}$, $\overline{v^{\prime 2}}$, $\overline{w^{\prime 2}}$ and $\overline{u^{\prime} v^{\prime}}$ at different downstream locations $x_p$ defined in Table \ref{tab:proflocs} are shown in Figures \ref{fig:uu_vertprofs}--\ref{fig:uv_vertprofs}. As was the case for the transverse profiles of the mean velocity components in Figures \ref{fig:u_vertprofs}--\ref{fig:v_vertprofs}, we note the similarity between the flows DNS400, EX12K and EX20K, and the distinct behaviour exhibited for DNS150.

The coherent contribution due to vortex shedding is expected to be large. Since the periodic shedding frequency and the drag coefficients are similar for all cases, the strength of shed vortices is also expected to be similar~\cite{roshko1954}, such that in the vortex-formation region close to the leeward side of the plate the fluctuation energy, defined as the turbulent kinetic energy $k=\frac{1}{2} ( \overline{u'^2}+\overline{v'^2}+\overline{w'^2} )$, is expected to be similar for all cases, with $\overline{v'^2}$ dominating. The redistribution of the fluctuation to the remaining Reynolds stresses is expected to occur through increased 3D instantaneous motion while lateral diffusion is expected to occur due to increased incoherent contribution. The production of $\overline{u'v'}$ is expected to be mainly due to the 2D deformation of the periodic vortices and thus change little with $Re$.  The present results are consistent with weaker 3D motion and lower incoherent contributions for $Re=150$ compared to the other cases. In particular, for $Re=150$, $\overline{u'^2}(x_p,0,0)$ and $\overline{w'^2}(x_p,0,0)$ are lower, while $\overline{v'^2}(x_p,0,0)$ higher, than for the other cases. Moreover, lateral diffusion of the Reynolds stresses is significantly reduced. Finally, the difference between $Re=150$ and the other cases would be expected to increase downstream of the formation region as the incoherent contribution to the Reynolds stress production increases.

\FloatBarrier

\subsection{Energy Budgets}\label{sub:budgets}

\begin{figure}[H]
    \centering
    \mbox{
        \subfigure[]
        {\includegraphics[scale=0.5]{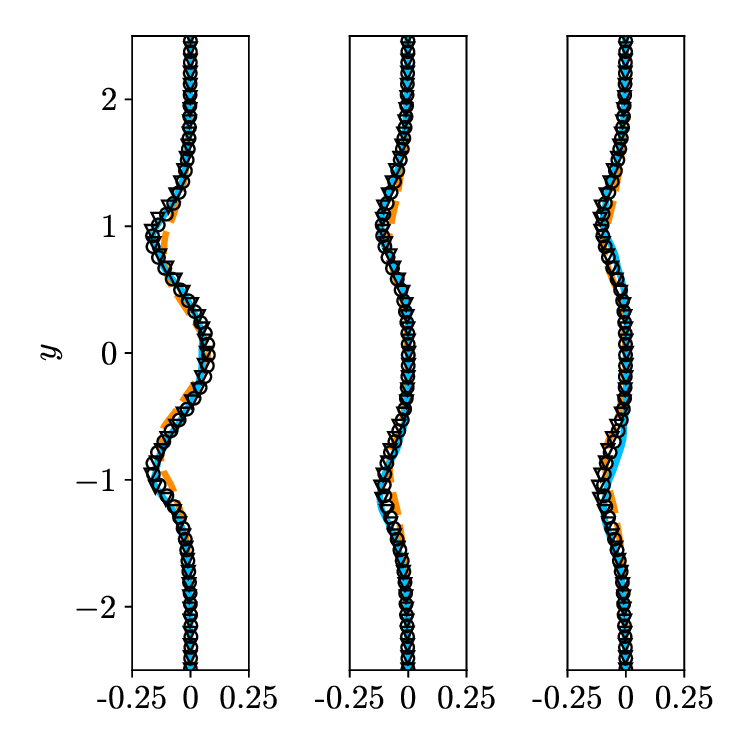}}
        \subfigure[]
        {\includegraphics[scale=0.5]{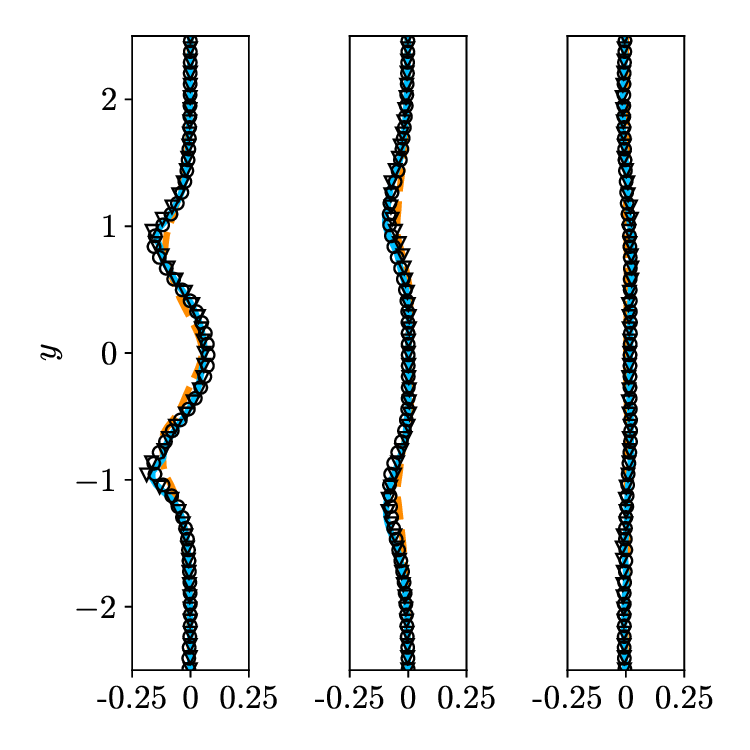}}}
\caption{Transverse profiles of the advection term in the $k$-transport equation \eqref{eq:ktrans} (a) at the downstream locations $x_p$ where the maxima of the Reynolds stresses (from left to right) $\overline{u^{\prime 2}}$, $\overline{u^\prime v^\prime}$ and $\overline{v^{\prime 2}}$ are attained, and (b) at (from left to right) $x_p = \ell/2$, $x_p = \ell$ and $x_p = 4.5$, cf.~Table \ref{tab:proflocs}. Legend: DNS150 \Plot{orange,dashed}; DNS400 \Plot{cyan,solid}; EX12K ($\bigcirc$); EX20K ($\bigtriangledown$).}
\label{fig:advec_vertprofs}
\end{figure}

\begin{figure}[H]
    \centering
    \mbox{
        \subfigure[]
        {\includegraphics[scale=0.5]{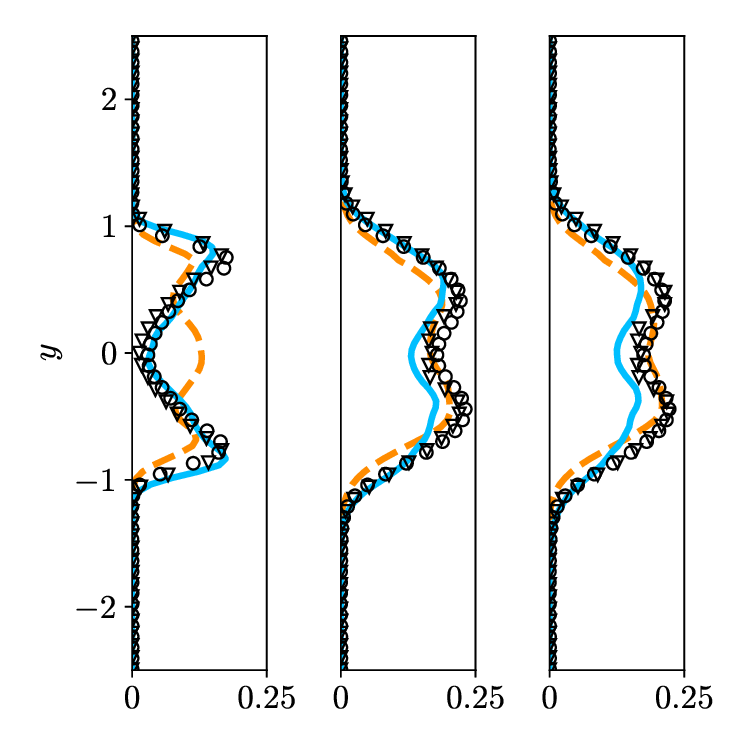}}
        \subfigure[]
        {\includegraphics[scale=0.5]{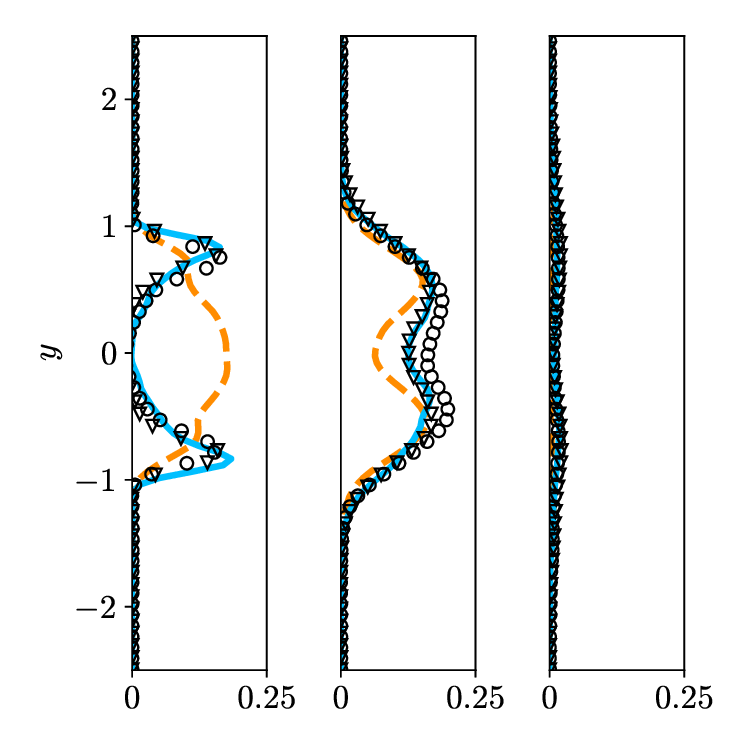}}}
\caption{Transverse profiles of the turbulence production term in the $k$-transport equation \eqref{eq:ktrans} (a) at the downstream locations $x_p$ where the maxima of the Reynolds stresses (from left to right) $\overline{u^{\prime 2}}$, $\overline{u^\prime v^\prime}$ and $\overline{v^{\prime 2}}$ are attained, and (b) at (from left to right) $x_p = \ell/2$, $x_p = \ell$ and $x_p = 4.5$, cf.~Table \ref{tab:proflocs}. Legend: DNS150 \Plot{orange,dashed}; DNS400 \Plot{cyan,solid}; EX12K ($\bigcirc$); EX20K ($\bigtriangledown$).}
\label{fig:produc_vertprofs}
\end{figure}

In order to provide insights on similarities and differences in the physical mechanism underlying turbulent transport in the considered flows, in this section we analyze the pointwise balance of the turbulent kinetic energy given by the $k$-transport equation for statistically stationary flows:
\begin{multline}
    -\underbrace{U_j\frac{\partial k}{\partial x_j}}_\text{Advection} + \underbrace{-\frac{1}{\rho}\frac{\partial \overline{u_i^\prime p^\prime}}{\partial x_i}}_\text{Pressure Diffusion} + \underbrace{-\frac{1}{2}\frac{\partial \overline{u^\prime_j u^\prime_j u_i^\prime}}{\partial x_i}}_\text{Turbulent Transport} + \underbrace{\nu\frac{\partial^2 k}{\partial x_j \partial x_j}}_\text{Molecular Diffusion} \\
    \underbrace{-\overline{u_i^\prime u_j^\prime}\frac{\partial U_i}{\partial x_j}}_\text{Turbulence Production} - \underbrace{\nu\overline{\frac{\partial u_i^\prime}{\partial x_j}\frac{\partial u_i^\prime}{\partial x_j}}}_\text{Dissipation, $\epsilon$} = 0,\label{eq:ktrans}
\end{multline}
where $U_j$ represents the mean velocity component (denoted $U$, $V$, or $W$ elsewhere). 
We focus on advection and turbulence production.
Their transverse profiles are compared at downstream locations $x_p$, defined in Table~\ref{tab:proflocs} and Figure~\ref{fig:probelocs}, in Figures \ref{fig:advec_vertprofs}--\ref{fig:produc_vertprofs}. 
For the experimental data, it was verified that $W$, $\overline{u'w'}$, $\overline{v'w'}$ and spanwise gradients vanish within experimental uncertainty. Hence, the $2D$ condition is enforced for all quantities. The remaining terms of (\ref{eq:ktrans}) are omitted for brevity. The turbulent transport shows similar trends to the previous terms without adding further insight, while the molecular diffusion terms was verified to be negligibly small as expected for higher $Re$. The pressure diffusion and dissipation $\epsilon$ could not be measured and thus comparison is not possible. 

As observed for the quantities discussed in \S~\ref{sub:meanvels}, a good agreement is evident between the distribution of these terms in the cases DNS400 and EX12K, EX20K, with the flow DNS150 revealing a distinct behaviour, especially with respect to the turbulence production.
The production terms arise due to nonlinear interactions resulting from internal deformation and Reynolds stress anisotropy. Hence, these results suggest that the flow dynamics are similar for $Re \ge 400$ for the different scales of motions, whereas for $Re=150$, the interactions between some motions are expected to differ. These observations would be consistent with differences in the fluctuation spectra and probability distributions that are discussed next.

\FloatBarrier
\subsection{Energy Spectra}\label{sub:spects}

\begin{figure}[H]
    \centering
    \mbox{
        \subfigure[]
        {\includegraphics[scale=0.3]{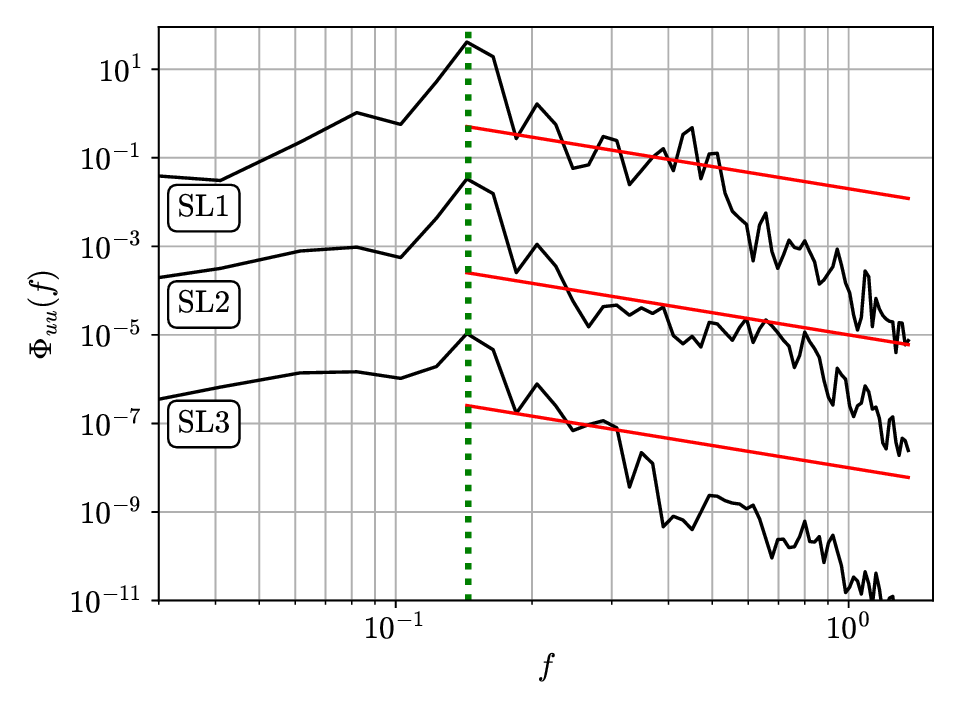}}
        \subfigure[]
        {\includegraphics[scale=0.3]{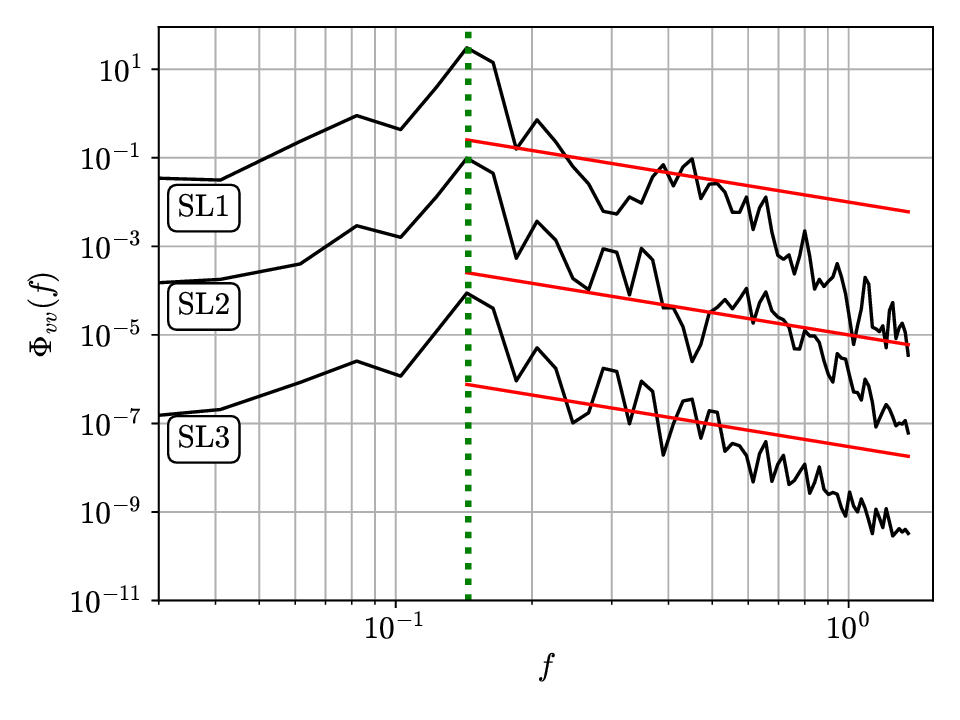}}
        \subfigure[]
        {\includegraphics[scale=0.3]{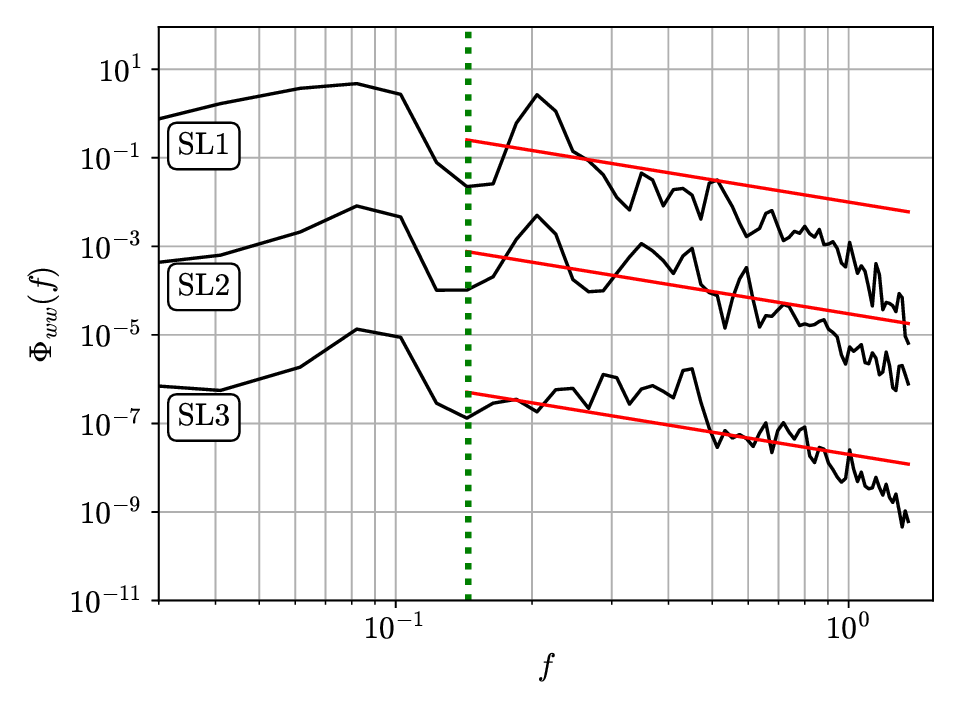}}}
    \mbox{
        \subfigure[]
        {\includegraphics[scale=0.3]{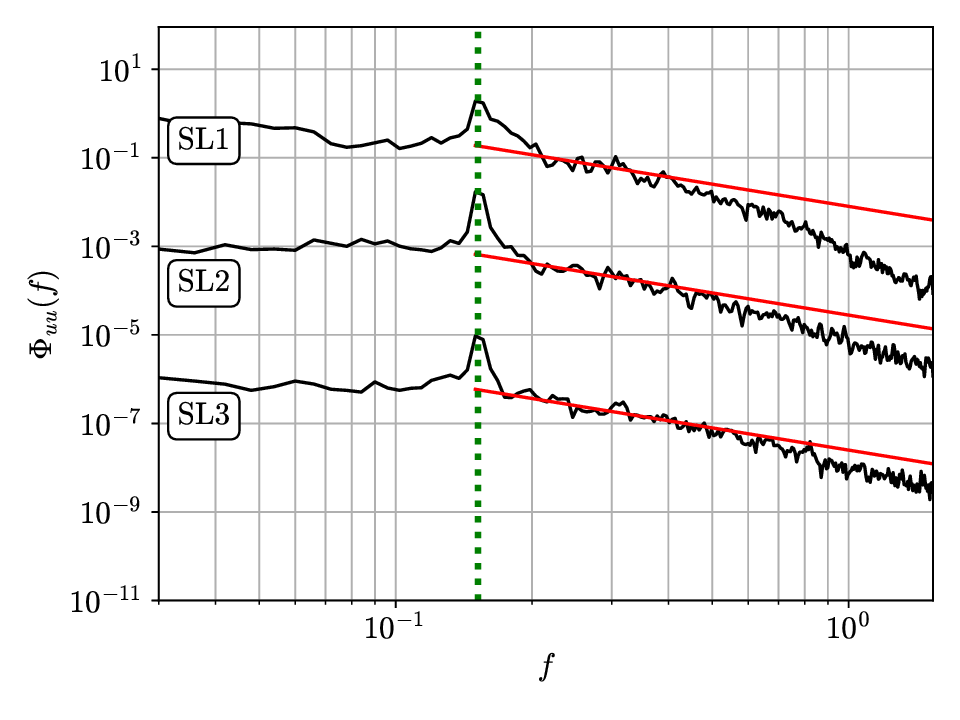}}
        \subfigure[]
        {\includegraphics[scale=0.3]{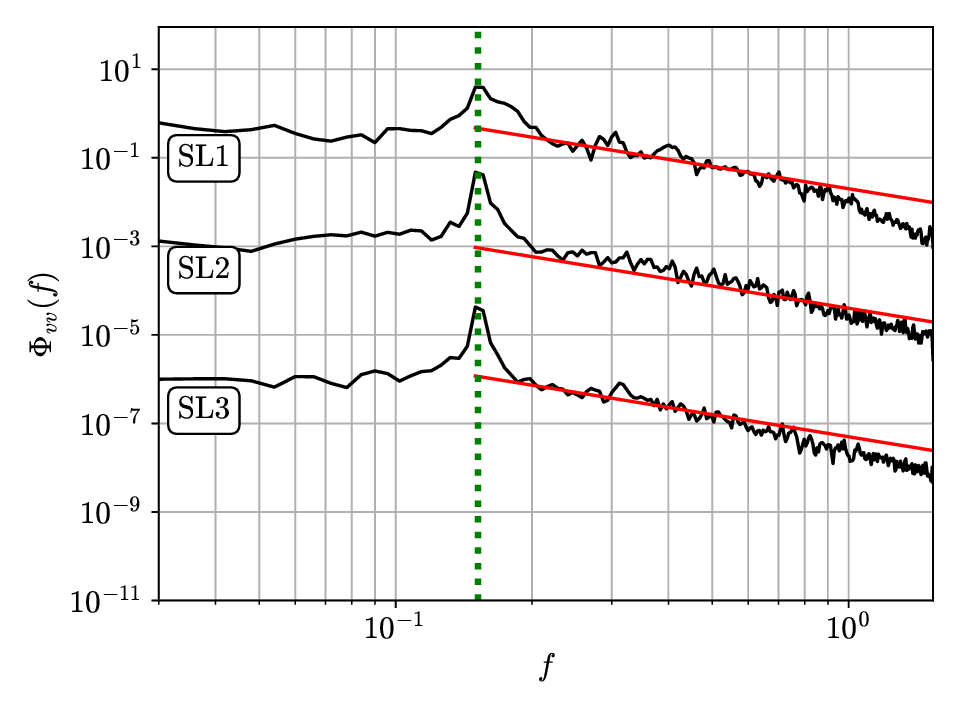}}
        \subfigure[]
        {\includegraphics[scale=0.3]{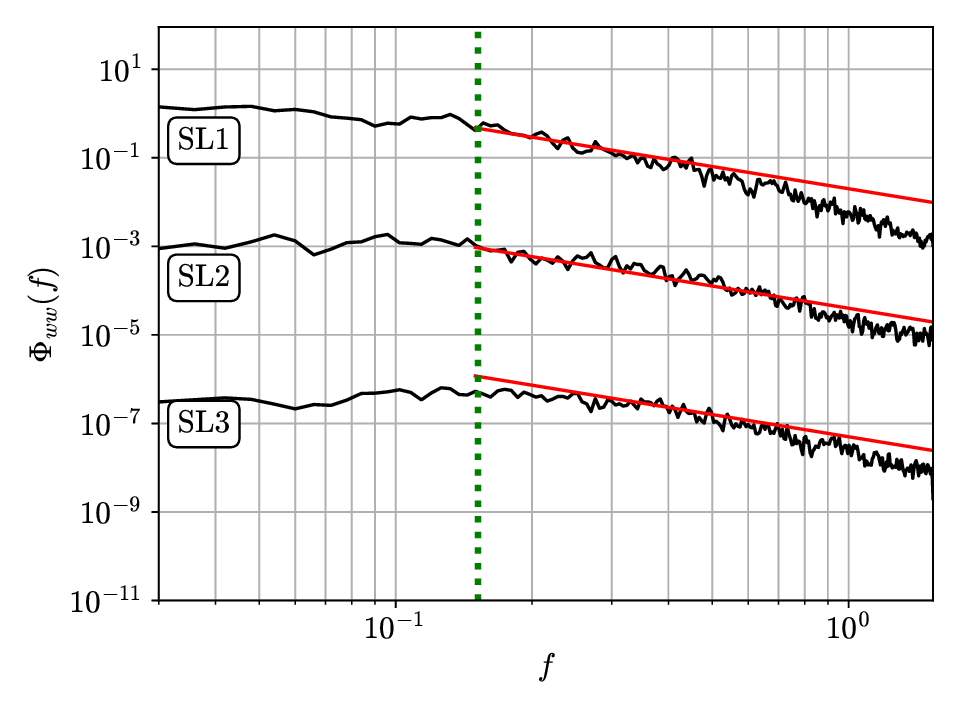}}}
    \mbox{
        \subfigure[]
        {\includegraphics[scale=0.3]{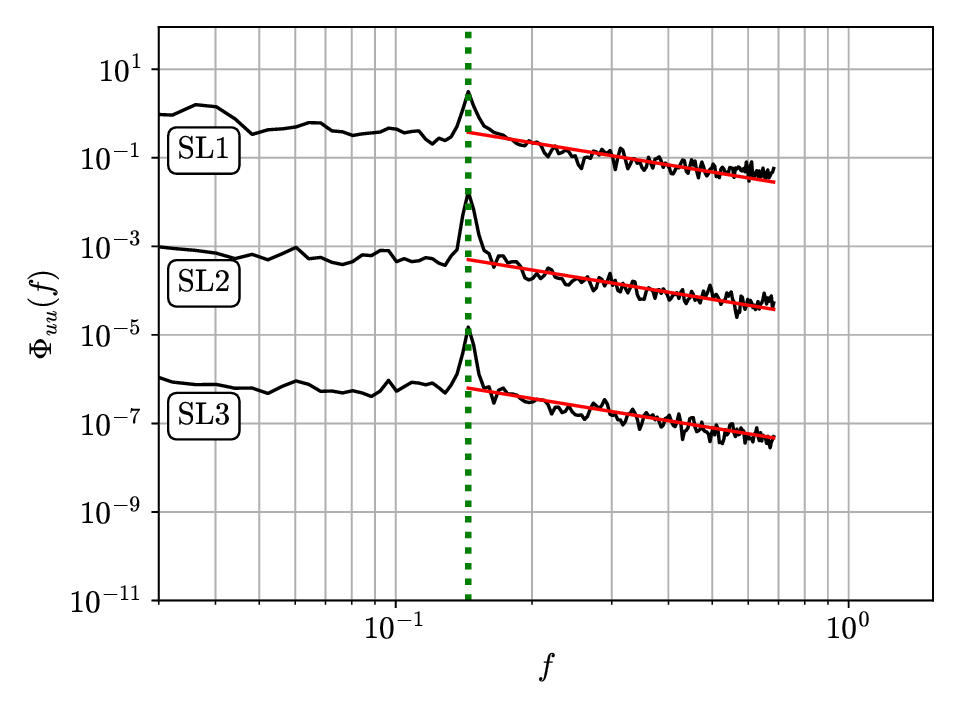}}
        \subfigure[]
        {\includegraphics[scale=0.3]{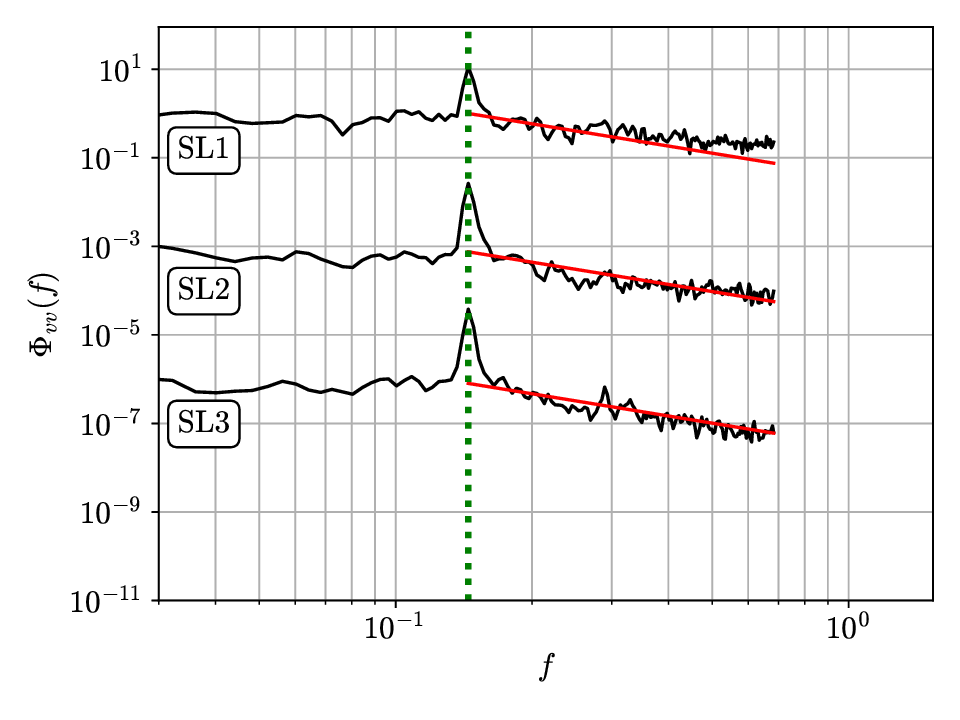}}
        \subfigure[]
        {\includegraphics[scale=0.3]{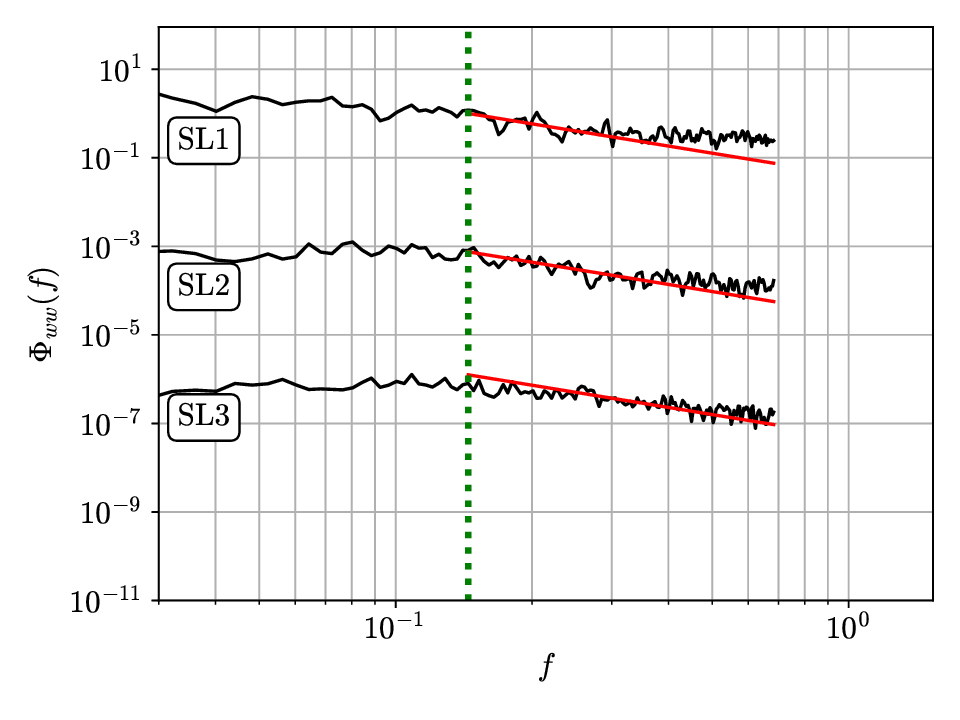}}}
\caption{Energy spectra of the velocity components (a) $u$, (b) $v$, and (c) $w$ in the flow DNS150, (d) $u$, (e) $v$, and (f) $w$ in the flow DNS400, and (g) $u$, (h) $v$, and (i) $w$ in the flow EK12K at the  locations in the shear layer indicated in Figure \ref{fig:probelocs}, see also Table \ref{tab:probelocs2} (the spectra corresponding to these different locations are offset vertically by three decades). The red lines represent the expression $C f^{-5/3}$ with different $C>0$, whereas the dotted vertical lines correspond to the normalized vortex-shedding frequency $St$ in this flow.}
\label{fig:shear_spects}
\end{figure}

The turbulent character of the wake flow at $Re=400$ (DNS400) is evident when comparing the power spectral density functions, $\Phi_{uu}, \Phi_{vv}$ and $\Phi_{ww}$ of the three velocity components to those from the experiment at $Re=12 500$ (EX12K) and the simulation at $Re=150$ (DNS150) as shown in Figure~\ref{fig:shear_spects}. The spectra are obtained by recording the velocity components $u(t)$, $v(t)$ and $w(t)$ as functions of time $t$ at three different locations in the shear layer indicated in Figure~\ref{fig:probelocs}, see also Table \ref{tab:probelocs2}, and then computing their discrete Fourier transforms $\widehat{u}(f)$, $\widehat{v}(f)$, $\widehat{w}(f)$, where $f$ is the frequency.
The dominant peaks in the spectra of $u$ and $v$ correspond to the vortex-shedding (Strouhal) frequency.
These peaks are absent from the energy spectra of $w$ since the wake flows are dominated by the von K{\'a}rm{\'a}n vortices associated with fluid motions mainly in the $x$--$y$ plane.

The spectra in Figure~\ref{fig:shear_spects} for $Re=400$ and $Re = 12500$ are very similar (noting that for the experiments the Kolmogorov scale is not resolved). These are clearly broad-band representing an energy cascade across different length- and time-scales, which is a hallmark of a turbulent flow. At each location an inertial range where the energy spectra scale approximately as $\Phi_{uu}, \Phi_{vv},\Phi_{ww} \sim f^\alpha$ with $\alpha <0$ can be discerned, although the extent of this range is rather narrow (less than a decade in $f$) which is expected for low $Re$.
In both cases in the vicinity of the plate the exponent $\alpha$ is less than Komogorov's K41 prediction of $\alpha = -5/3$. Further downstream, at SL2 and SL3, the spectra for all components approach the scaling with $\alpha = -5/3$ indicating an inertial subrange with local isotropy. In contrast, while the spectra for $Re=150$ show chaotic and 3D character, they do not reveal the multiscale behavior expected for turbulent flows. More specifically, these spectra show energy transfers between a small numbers of modes only.


\begin{table}[H]
    \centering
    \begin{tabular}{r|lll}\hline\hline
        \rule{0pt}{2.5ex}Probe Name & DNS150 & DNS400 & EX12K \\
        \hline
        SL1 & (0.505,0.732) & (0.487,0.789) & (0.537,0.669) \\
        SL2 & (1.364, 0.631) & (1.343,0.856) & (1.220,0.754) \\
        SL3 & (2.222, 0.732) & (2.200,0.554) & (1.989,0.583) \\\hline\hline
    \end{tabular}
    \caption{Locations of the shear layer (SL) probes in the plane $z=0$ used in DNS150, DNS400 and EX12K to produce the spectra shown in Figure \ref{fig:shear_spects}. The probe locations in DNS400 are illustrated in Figure \ref{fig:probelocs}.}
    \label{tab:probelocs2}
\end{table}

\FloatBarrier

\subsection{Non-Gaussianity of Turbulence Statistics}\label{sub:hist}

A salient feature of turbulence is non-Gaussian distribution of flow quantities involving spatial derivatives, such as vorticity and local energy dissipation \cite{frisch1995turbulence}. Here we illustrate this aspect for the flow DNS400 in contrast to DNS150. Following Buaria et al. \cite{buaria2019}, we define the quantities
\begin{align}
    \Omega & = \omega'_i\omega'_i,  \label{eq:Omega} \\
    \Sigma & = 2s'_{ij}s'_{ij}, \label{eq:Sigma} \\
    \tau_K & = \sqrt{\frac{\nu}{\epsilon}},
\end{align}
where $\omega'_j$ is the $j$th component of the fluctuating vorticity $\boldsymbol{\omega}' = \nabla \times \mathbf{u}'$, $s'_{ij}=(\partial u'_i/\partial x_j + \partial u'_j/\partial x_i)/2$  the $ij$th entry of the fluctuating strain rate tensor, and $\tau_K$ the Kolmogorov time scale.
The quantities $\Omega$ and $\Sigma$ in \eqref{eq:Omega}--\eqref{eq:Sigma} are functions of both time and space, and 
upon integration over the flow domain (and multiplication by $\nu$ in the latter case), they give, respectively, the total instantaneous enstrophy of fluctuations and energy dissipation.

\begin{figure}
    \centering
    \mbox{
        \subfigure[]
        {\includegraphics[scale=0.5]{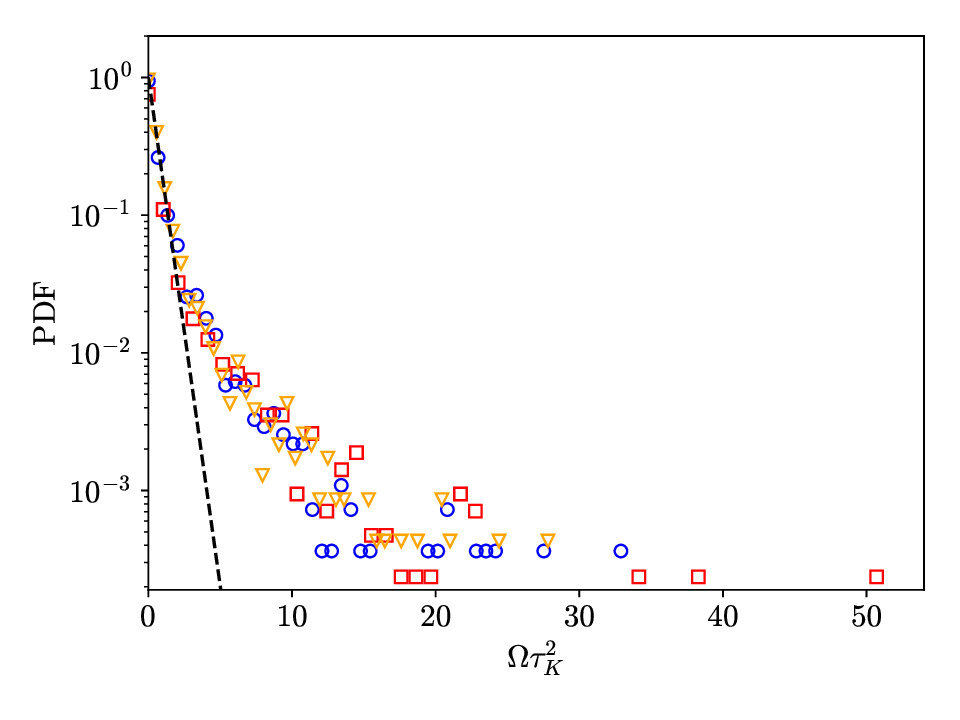}}
        \subfigure[]
        {\includegraphics[scale=0.5]{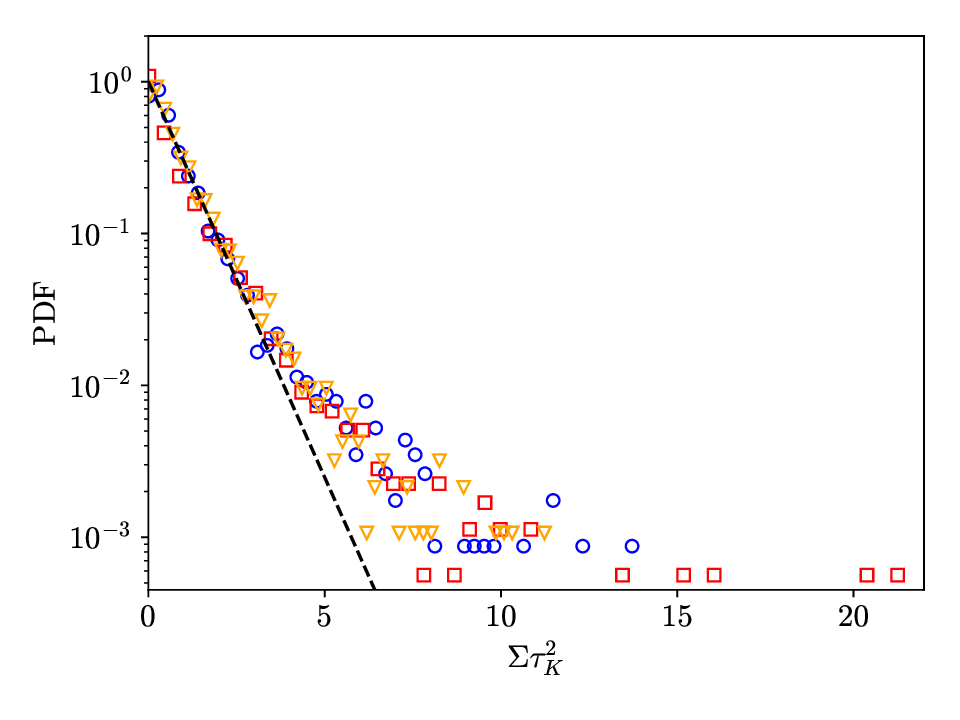}}}
\caption{PDFs for DNS400 of (a) $\Omega\tau_K^2$ and (b) $\Sigma\tau_K^2$ recorded at at the downstream locations $x_p$ where the maxima of the Reynolds stresses $\overline{u^{\prime 2}}$ (\color{blue}$\bigcirc$\color{black}), $\overline{v^{\prime 2}}$ (\color{red}$\square$\color{black}), and $\overline{w^{\prime 2}}$ (\color{orange}$\bigtriangledown$\color{black}) are attained. The dashed lines represent the Gaussian distributions $\exp(-\beta \Omega \tau_K^2)$ and $\exp(-\beta \Sigma \tau_K^2)$ with arbitrary parameters $\beta$.}
\label{fig:Re400_hists}
\end{figure}

\begin{figure}
    \centering
    \mbox{
        \subfigure[]
        {\includegraphics[scale=0.5]{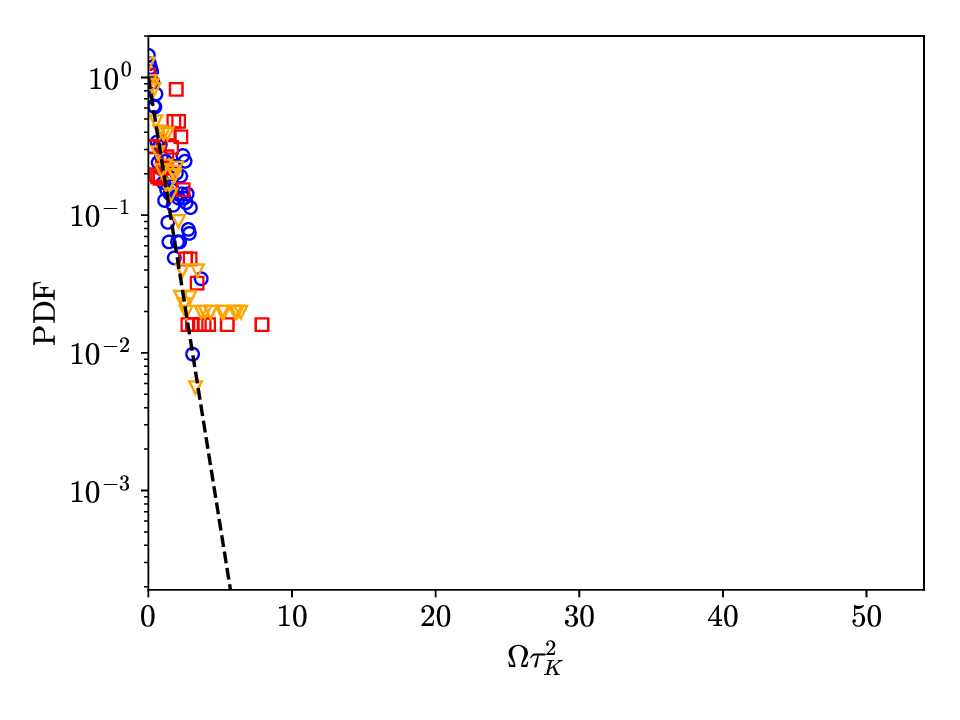}}
        \subfigure[]
        {\includegraphics[scale=0.5]{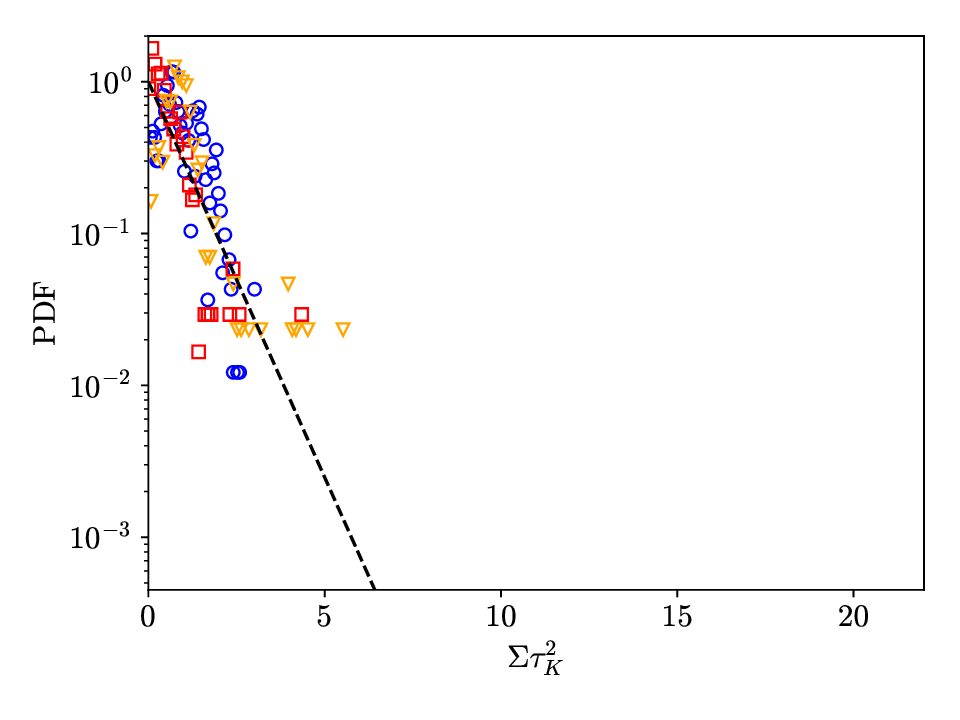}}}
\caption{PDFs for DNS150 of (a) $\Omega\tau_K^2$ and (b) $\Sigma\tau_K^2$ recorded at at the downstream locations $x_p$ where the maxima of the Reynolds stresses $\overline{u^{\prime 2}}$ (\color{blue}$\bigcirc$\color{black}), $\overline{v^{\prime 2}}$ (\color{red}$\square$\color{black}), and $\overline{w^{\prime 2}}$ (\color{orange}$\bigtriangledown$\color{black}) are attained. The dashed lines represent the Gaussian distributions $\exp(-\beta \Omega \tau_K^2)$ and $\exp(-\beta \Sigma \tau_K^2)$ with arbitrary parameters $\beta$.}
\label{fig:Re150_hists}
\end{figure}

The probability density functions (PDFs) of $\Omega$ and $\Sigma$  normalized with $\tau_K^2$ and recorded at the locations of the maxima of the normal Reynolds  
stresses (i.e., $\overline{u^{\prime 2}}$, $\overline{v^{\prime 2}}$ and $\overline{w^{\prime 2}}$, {\it cf.}~Figure \ref{fig:probelocs}) are shown in Figure \ref{fig:Re400_hists} and \ref{fig:Re150_hists} for $Re=400$ and 150, respectively. At each of these locations, the PDFs are approximated as histograms of the time series of $\Sigma \tau_K^2$ and $\Omega \tau_K^2$. If the different velocity derivatives were to follow Gaussian distributions, i.e., if for some $\xi = \pdv{u'_i}{x_j}$ we would have $PDF(\xi) \sim \exp(-\beta \xi^2)$ with a certain $\beta >0$, then their squares $\eta = \xi^2$ would follow exponential distributions $PDF(\eta) \sim \exp(-\beta \eta)$, $\eta \ge 0$, represented in Figure \ref{fig:Re400_hists} and \ref{fig:Re150_hists} as straight lines with the slope $-\beta$. For $Re=400$, the quantities $\Omega$ and $\Sigma$ exhibit at all three locations quite different distributions characterized by heavy ``tails" at sufficiently large values. These tails correspond to rare high-intensity events, involving concentrated vorticity for $\Omega$ and localized dissipation for $\Sigma$, which result from the intermittent nature of turbulent flows \cite{buaria2019,frisch1995turbulence}. In contrast, these characteristic features of turbulence are not observed for $Re=150$. This thus provides yet another demonstration that the wake flow DNS400 is indeed turbulent.

\FloatBarrier
\subsection{Properties of the Far Wake}\label{sub:farwake}

\begin{figure}[H]
    \centering
    \vspace{-2cm}
    \subfigure[]
    {\includegraphics[scale=0.4]{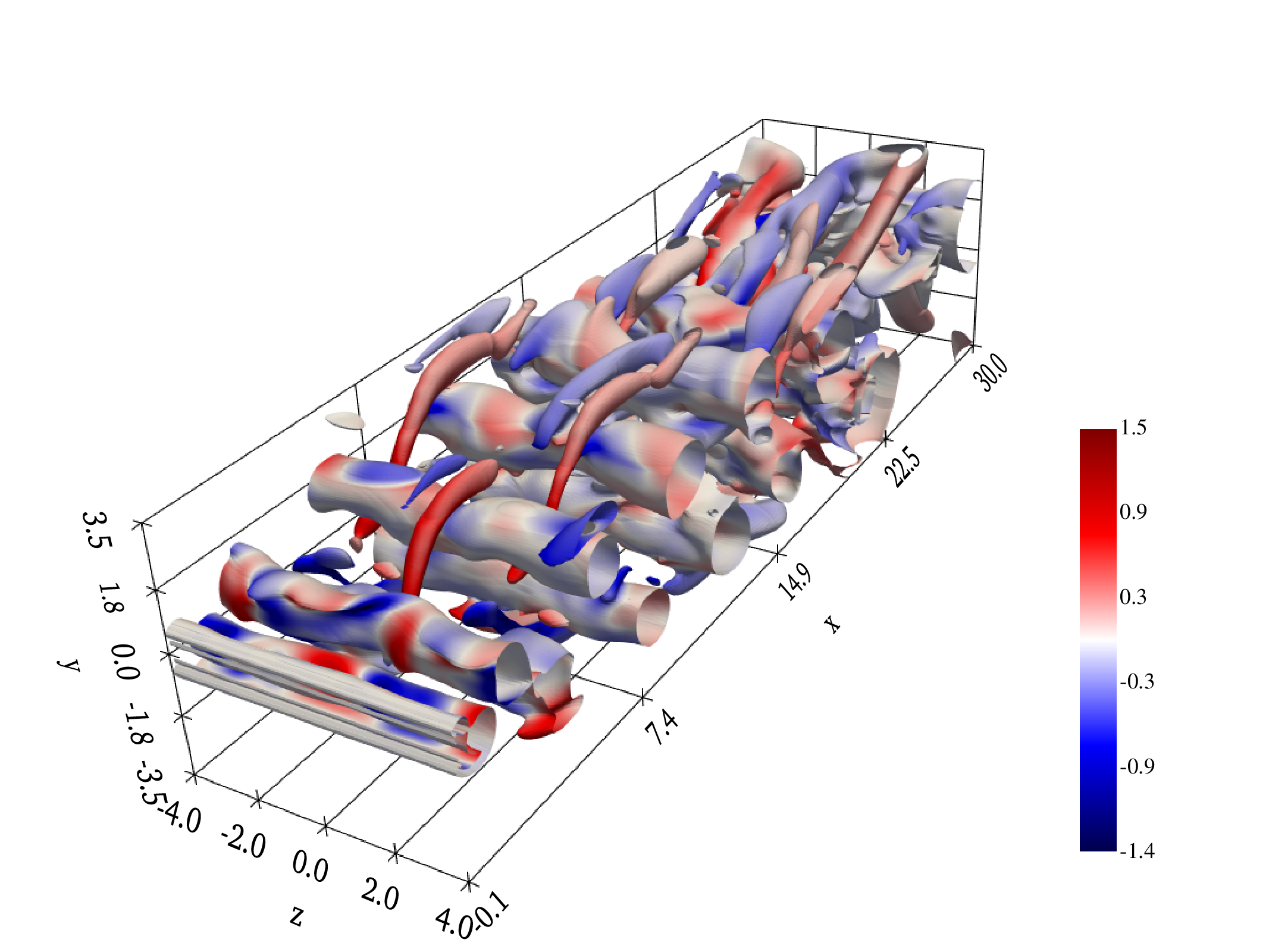}}
    \subfigure[]
    {\includegraphics[scale=0.4]{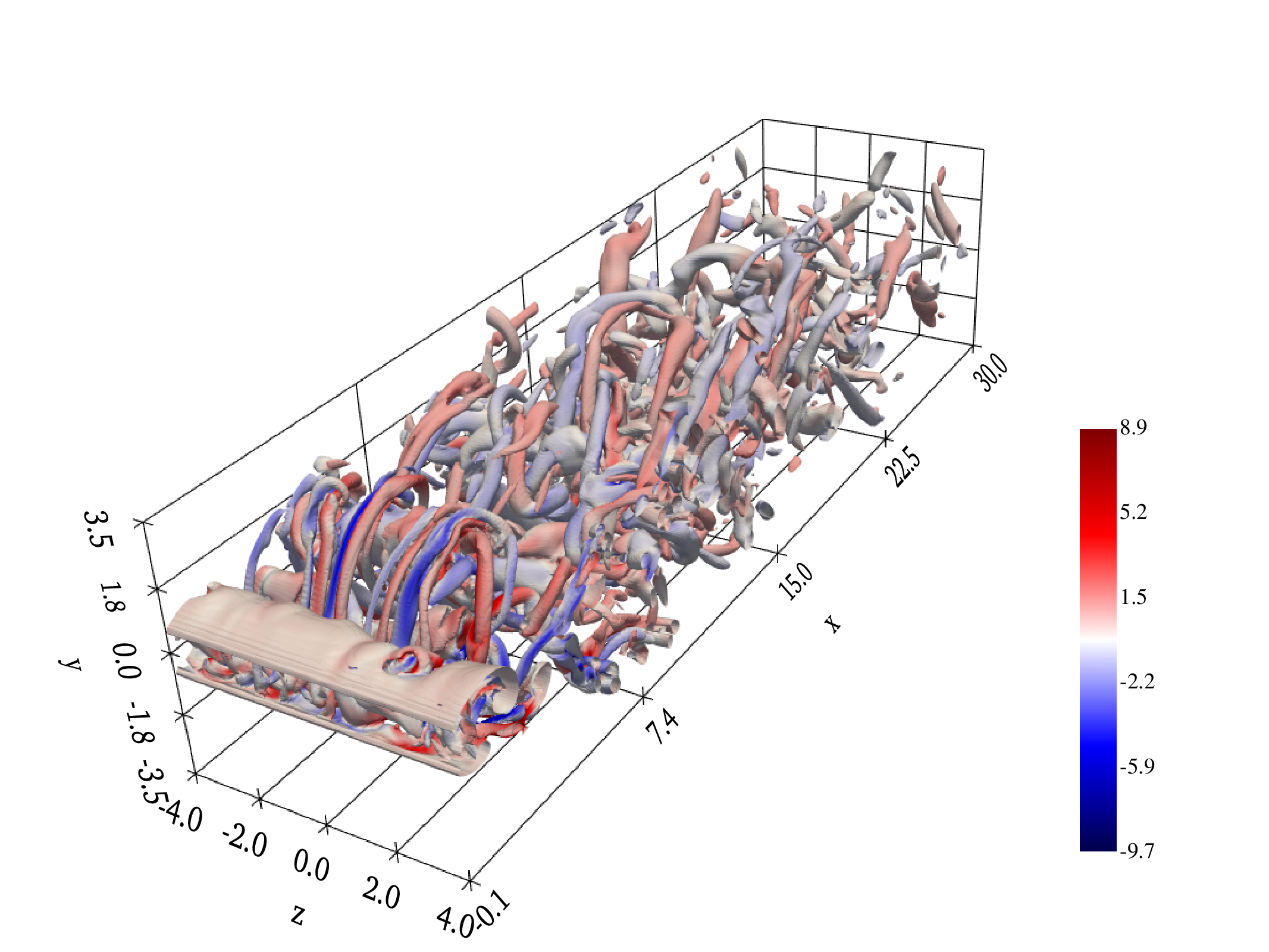}}
\caption{Isosurfaces of the second invariant $Q$ of the instantaneous strain rate tensor coloured with the values of the longitudical vorticity component $\omega_1$ in (a) DNS150 with $Q(x,y,z)=0.001$ and (b) DNS400 with $Q(x,y,z)=0.2$. Colour bars show the values of $\omega_1$ and correspond to the entire ranges of vorticity values in the flow fields.}
\label{fig:Sausages}
\end{figure}

We close the discussion by considering the behaviour of the wake flows DNS150 and DNS400 far downstream from the obstacle. The instantaneous vortical structures in the unsteady flows at $Re=150$ and $Re=400$ are illustrated at arbitrary times in Figure~\ref{fig:Sausages} using iso-surfaces of the second invariant 
$Q= \frac{1}{2}\frac{\partial u_i}{\partial x_j} \frac{\partial u_j}{\partial x_i}$ 
of the instantaneous strain rate tensor to identify the vortex cores. 
The iso-surfaces $Q(x,y,z)$ are colour-coded by the streamwise vorticity component.
The figure shows that the path leading to transition differs for the thin flat plate compared to the canonical cylinder geometries. For $Re=150$, while 3D instabilities arise in the near wake, these do not give rise to coherent structures other than the von K\'{a}rm\'{a}n rollers. The rollers are only weakly distorted and remain coherent at least 5$d$ downstream of the plate. Further downstream, additional instabilities appear that alter the structure of the flow. An analogous regime is not observed for circular or square cylinder wakes even immediately after the onset of 3D instabilities. At $Re=400$, the flow exhibits a highly 3D character, replete with streamwise ribs and strongly distorted rollers apparent in the near wake with  distortion increasing downstream. These effects result in the formation of fine flow structures already close  to the obstacle. The transition between these two wake states occurs over a very short $Re$ range and does not have an analog for flows past canonical cylinders.

Planar wake flows, both laminar and turbulent, are known to exhibit a self-similar behaviour far downstream from the obstacle which can be characterized in terms of the wake deficit defined as
\begin{equation}
\Delta U(x,y) = 1 - \frac{U(x,y)}{U_\infty}.
\label{eq:dU}
\end{equation}
The classical similarity solution for such flows posits the existence of a universal profile $F(\eta)$ attained asymptotically at large downstream distances from the obstacle  \cite{schlichting2016}
\begin{equation}
\Phi(x,\eta) := 
\frac{\Delta U(x,\eta)}{\Delta U(x,0)} \underset{x \rightarrow \infty} \longrightarrow  e^{-\eta^2} =: F(\eta) \quad \text{where} \quad  \eta = \sqrt{\ln 2} \frac{y}{y_{0.5}}
\label{eq:F}
\end{equation}
in which $y_{0.5}$ is the half-width of the wake, {\it i.e.}, the $y$ location where $F(\eta)=0.5$.  Relation \eqref{eq:F} stipulates that far downstream in a planar wake the mean longitudinal velocity $U$ approaches a Gaussian profile as a function of the suitably rescaled transverse coordinate $\eta$. The difference between the laminar and turbulent regimes is encoded in the spread rate of the Gaussian profile (i.e., $y_{0.5}$ in \eqref{eq:F}) and in how rapidly this asymptotic profile is approached \cite[Sections 7.5.1 and 22.5]{schlichting2016}. Moreover, this  theory  also predicts that
\begin{equation}
y_{0.5} \sim x^{1/2} \qquad \text{and} \qquad \Delta U(x,0) \sim x^{-1/2} \qquad  \text{as} \qquad x \rightarrow \infty,
\label{eq:farwake}
\end{equation}
meaning that the wake has a parabolic shape and becomes ``shallower" far downstream.

\begin{figure}[H]
    \centering
    \includegraphics[scale=0.5]{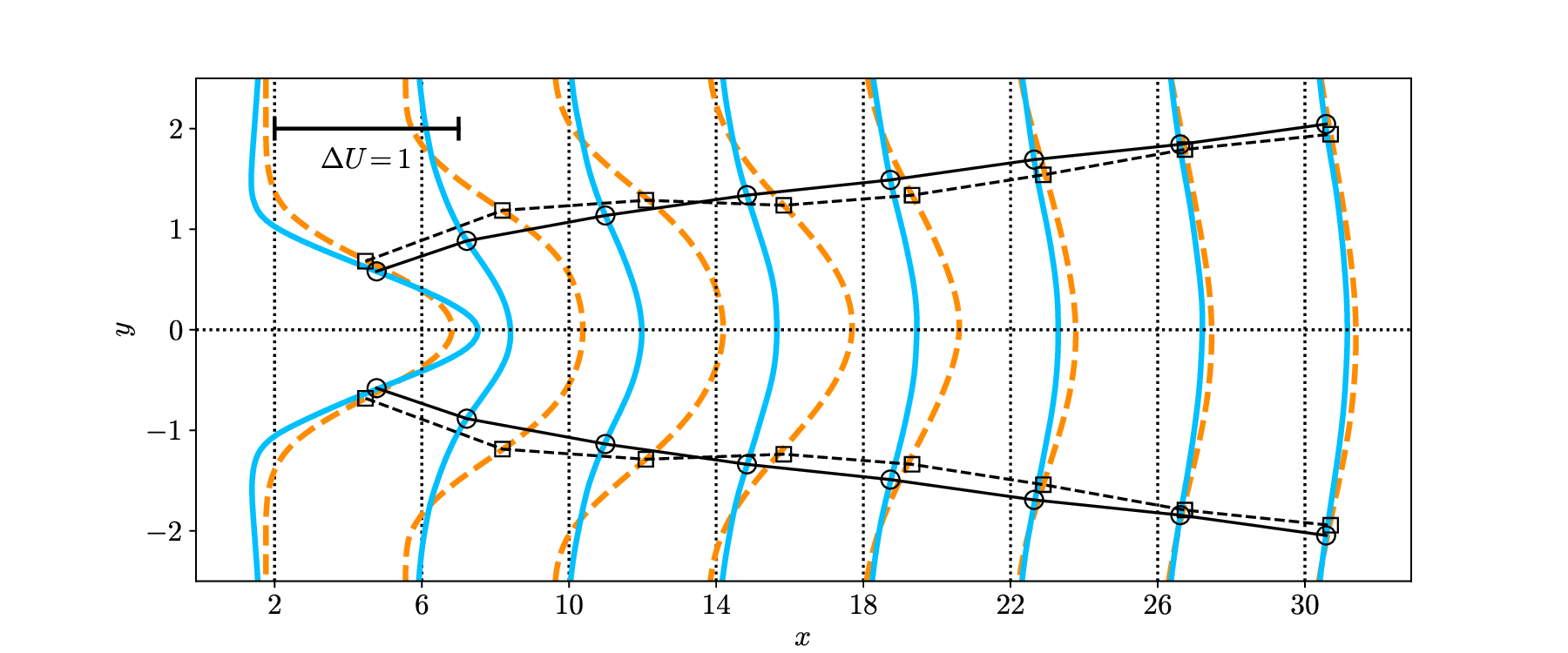}
    \caption{The wake deficits  $\Delta U(x,y)$ at the downstream locations indicated on the horizontal axis. Legend: DNS150 \Plot{orange,dashed}; DNS400 \Plot{cyan,solid}. The symbols and the straight lines represent the values of the wake half-width in the two cases, whereas the dotted vertical lines mark the origin at each value of $x$.}
    \label{fig:WakeSpreading}
\end{figure}

\begin{figure}[H]
    \centering
    \subfigure[]
    {\includegraphics[scale=0.4]{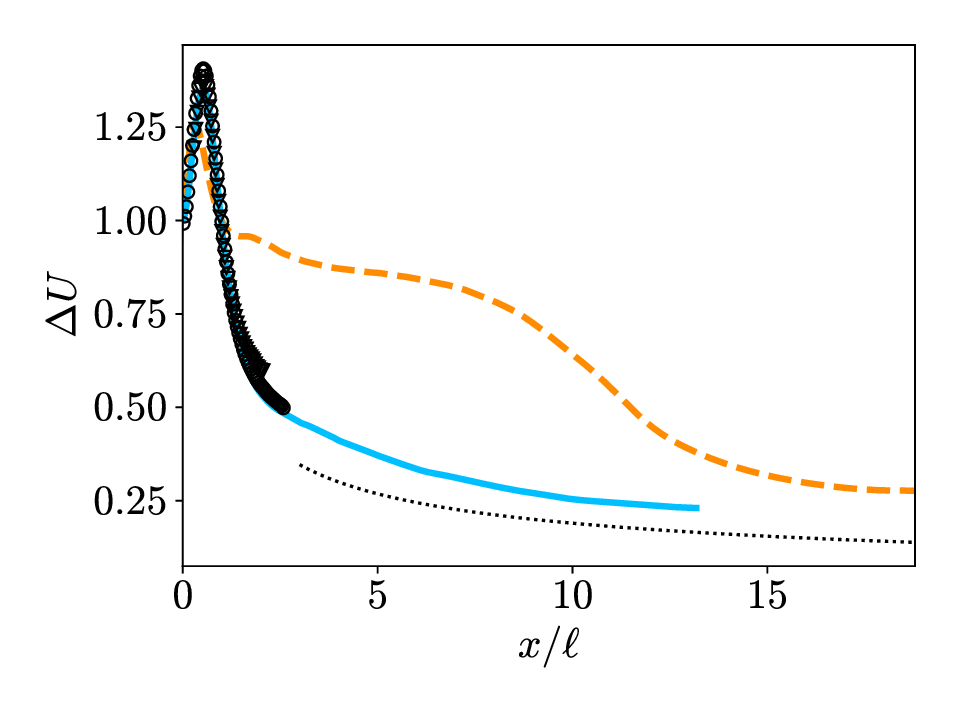}}
    \subfigure[]
    {\includegraphics[scale=0.4]{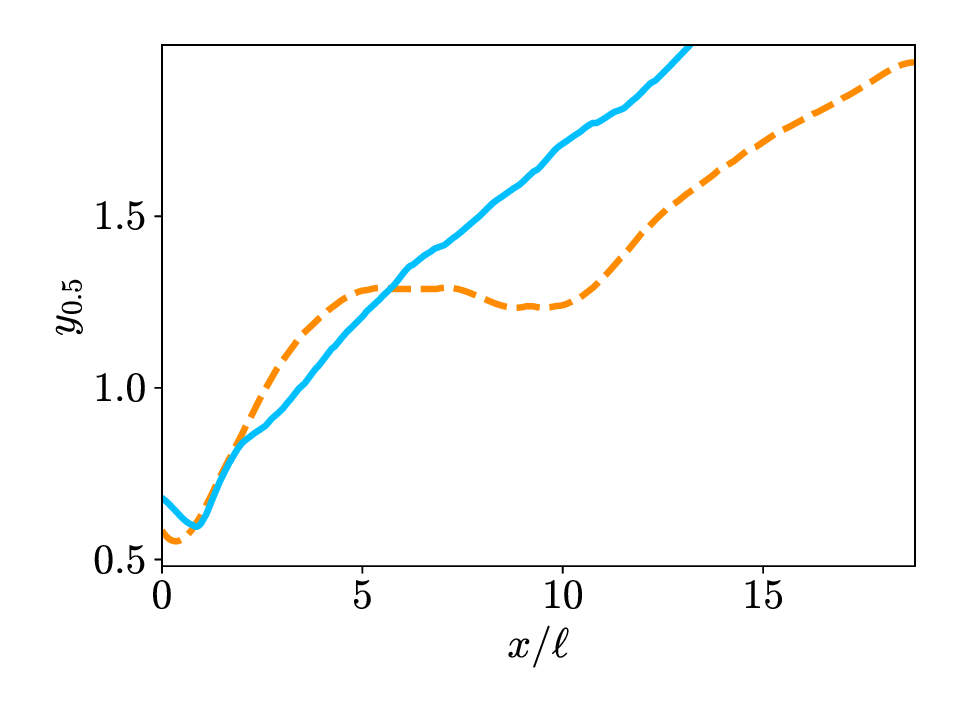}}
    \caption{The dependence of (a) the wake deficit $\Delta U$ and (b) the wake half width $y_{0.5}$ on $x/\ell$ along the flow centerline. Legend: DNS150 \Plot{orange,dashed}; DNS400 \Plot{cyan,solid}; EX12K ($\bigcirc$); EX20K ($\bigtriangledown$); relations \eqref{eq:farwake} with arbitrary prefactors and origins \Plot{black,dotted}.}
    \label{fig:wakeDefs}
\end{figure}

\begin{figure}[H]
    \centering
    \includegraphics[scale=0.5]{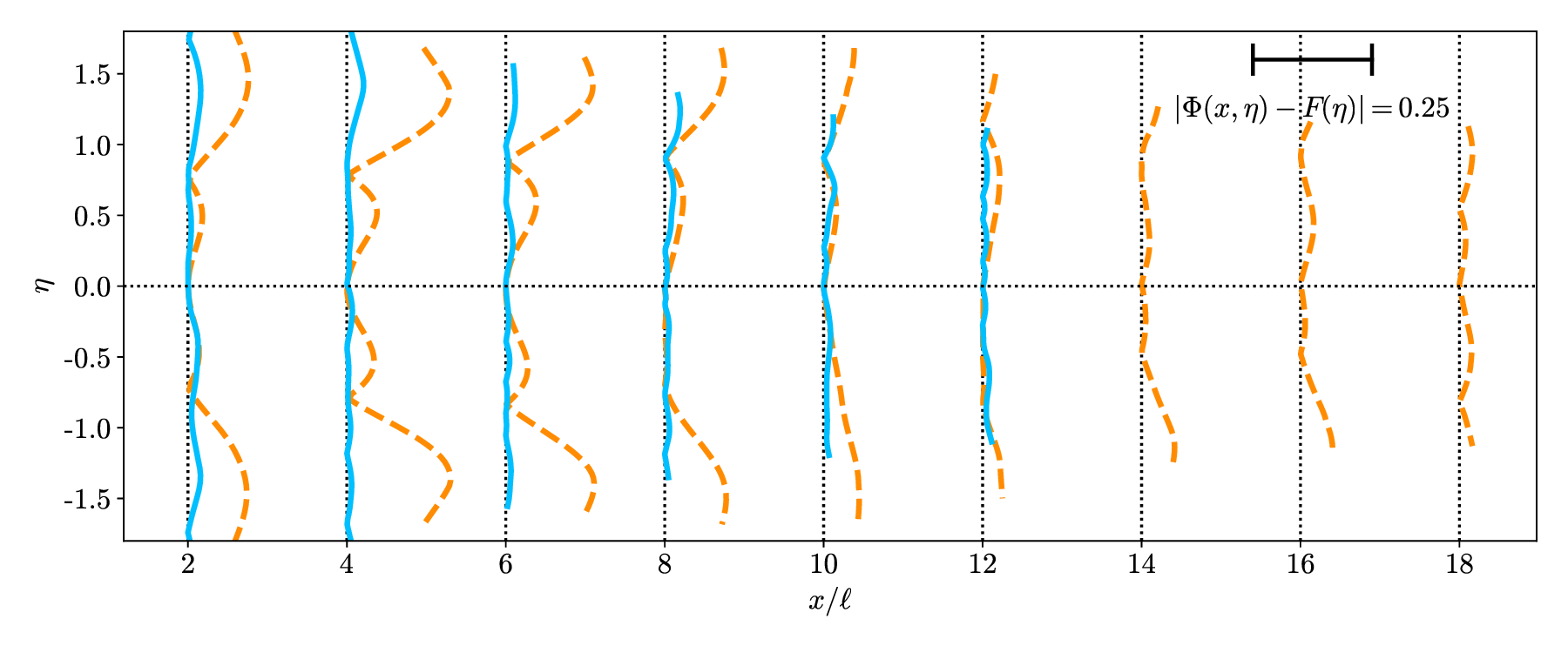}
    \caption{The difference $
    |\Phi(x,\eta) - F(\eta)|$ between the rescaled actual wake deficit and the theoretical prediction, cf.~\eqref{eq:F}, as function of $\eta$ at different normalized downstream locations $x / \ell$. Legend: DNS150 \Plot{orange,dashed}; DNS400 \Plot{cyan,solid}. The dotted vertical lines mark the origin at each value of $x / \ell$.}
    \label{fig:F_eta_diff}
\end{figure}

To gain insights about the behaviour of the wake deficit $\Delta U(x,y)$ at large downstream distances from the obstacle, the wake flows DNS150 and DNS400 are compared in Figure~\ref{fig:WakeSpreading} as a function of $x$ and $y$ at several increasing downstream locations. We see that for $x > 2$ the wake deficit near the flow centerline is noticeably smaller in the case DNS400. We also observe that in the case DNS150 the wake half-width $y_{0.5}$ exhibits a non-monotonic behaviour in contrast to the case DNS400. In order to quantify these observations and to assess how they compare with the predictions \eqref{eq:farwake}, in  Figure~\ref{fig:wakeDefs}a we show the evolution of the deficit $\Delta U(x/\ell,0)$ along the flow axis in all four flows considered here with the corresponding evolution of $y_{0.5}$ shown for the cases DNS150 and DNS400 in Figure~\ref{fig:wakeDefs}b. The wake deficit has a similar dependence on $x/\ell$ in the cases DNS400, EX12K and EX20K, and this behaviour is consistent with relation \eqref{eq:farwake}. On the other hand, in the case DNS150 the recovery of the deficit is much slower. In DNS400 the dependence of $y_{0.5}$ on $x/\ell$ is also consistent with \eqref{eq:farwake}, while for DNS150 $y_{0.5}$ exhibits a qualitatively different behaviour characterized by a non-monotonic dependence on $x/\ell$. More specifically, in this latter case, the wake half-width first increases, then decreases before finally approaching the asymptotic growth in $x/\ell$ predicted by \eqref{eq:farwake}.
The non-monotonic behaviour occurs in the region $5<x/\ell<10$, which corresponds to the region of increasingly complex three-dimensional structure of the DNS150 wake observed in $7<x<15$ in Figure~\ref{fig:Sausages}a. 
Finally, in Figure \ref{fig:F_eta_diff} we show
the difference $|\Phi(x,\eta) - F(\eta)|$ between the rescaled actual wake deficit and the theoretical prediction, cf.~\eqref{eq:F}, as a function of $\eta$ at different normalized downstream locations $x / \ell$ for the two cases. It is evident that in the wake flow DNS400 the approach to the asymptotic profile $F(\eta)$ is more uniform in $\eta$ and rapid than in DNS150, highlighting different physical mechanisms at play in the two flows.
    
\FloatBarrier

\section{Conclusions}\label{sec:conclusion}

The wake of a two-dimensional thin flat plate normal to a uniform stream was investigated using DNS at $Re=150$ and 400. Detailed comparison with results from experimental studies at $Re=12 500$ and $20 000$ provides evidence that the present wake is already fully turbulent at $Re=400$, which is lower than observed for the canonical circular and square cylinder wakes. The results at $Re=150$ moreover indicate that the transition to a turbulent wake follows a different path than for the canonical cases.

The distribution of the mean velocity, Reynolds stress and individual $k$-transport terms for the DNS at $Re=400$ and the experiments agree within experimental uncertainty. Comparison of the power spectral density functions of their fluctuating velocity components in the base and near wake regions shows similar spatial evolution and characteristic energy producing, inertial, and dissipative ranges. Probability density functions of the fluctuating strain and rotation rate magnitudes, $\Sigma$ and $\Omega$, show the non-Gaussian distribution of extreme events associated with intermittency which characterizes fully turbulent flows.

In contrast, at $Re=150$, the mean field differs significantly. Unlike in turbulent wakes, the spectra of the velocity do not show a full hierarchy of scales and the probability density functions of $\Sigma$ and $\Omega$ are Gaussian. The mean wake develops non-monotonically. Moreover, three-dimensional instability regimes are observed that do not have analogs in canonical cylinder wakes.

The present observations support the view that the low-$Re$ evolution of thin flat plate wakes differs fundamentally from that of canonical cylinder wakes. These observations also help reconcile some of the early results reported in the literature, especially related to the appearance of coherent wake structures and their sensitivity to the obstacle aspect ratio or $Re$. 
The physical mechanisms underlying these differences may be related to the influence of the fluctuating pressure field on the vorticity transport from the separation point. 
For obstacles with $b/d \gtrsim 0.6$ such as the circular and square cylinders, the streamwise extent of the obstacle ({\it i.e.,} the afterbody) has a stabilizing influence on the flow \cite{lynrodi1994}. In effect, the pressure fluctuations immediately downstream of the separation are significantly weaker than those arising in the vortex-formation region. In contrast, the pressure fluctuations are directly coupled to those in thin flat plate wakes \cite{braun2020}. 
The differences with respect to the transition scenarios in the canonical cylinder wakes reported here point to the need to describe more completely and, ultimately, to better understand the sequence of instabilities that lead to the early onset of turbulence in the wake past a thin flat plate.

\section*{Acknowledgements}

We thank the Natural Sciences and Engineering Research Council of Canada and the Digital Research Alliance of Canada for supporting this research. Scholarships from the Province of Alberta and the University of Calgary (Faculty of Graduate Studies) have partially supported Rosin.

\section*{Disclosure Statement}

We report no potential conflict of interest.

\section*{Data Availability Statement}

Data will be made available upon reasonable request.

\bibliographystyle{tfs}
\bibliography{bib}

\end{document}